\documentclass[]{aastex631}

\usepackage{amsmath}
\usepackage{comment}
\usepackage{booktabs}
\usepackage{xcolor}

\newcommand{\unit}[1]{\,\mathrm{#1}}
\newcommand{\e}[1]{\times 10^{#1}}
\providecommand{\Rsun}{R_\odot}

\newcommand{\RE}{R_\oplus}
\newcommand{\vsw}{v_{\mathrm{sw}}}
\newcommand{\Esep}{E_{\mathrm{SEP}}}
\newcommand{\fdep}{f_{\mathrm{dep}}}
\newcommand{\cell}[2]{\parbox[t]{#1\textwidth}{\linespread{1.0}\selectfont\raggedright\strut #2\strut}}

\shorttitle{Young-Sun CMEs: A Self-Extinguishing SEP Forcing}
\shortauthors{Panasenco}

\begin{document}

\title{Coronal Mass Ejections from the Young Sun: Deceleration-Powered Solar Energetic Particles \\ as a Self-Extinguishing Forcing of the Early Earth}

\author[0000-0002-4440-7166]{Olga A. Panasenco}
\affiliation{Advanced Heliophysics Inc., Pasadena, CA}

\begin{abstract}
At $\sim$3.9~Ga (stellar age $\sim$0.6~Gyr), the young Sun was far more magnetically active than
today. It launched $\sim$$10^2$ large, fast coronal mass ejections (CMEs) per year into a denser,
more magnetized wind. We follow these eruptions to Earth's atmosphere with a semi-analytic model
chain. We find that a CME launched at $v_0\sim3\text{--}4\e{3}\unit{km\,s^{-1}}$ dissipates
$\Delta E_k\sim6\e{25}\unit{J}$ as it decelerates, an amount set only by its launch and final
speeds. Diffusive shock acceleration converts $\sim$10\% of this energy into solar energetic
particles (SEPs), whose deposition depth depends mainly on a spectral break energy $E_b$. We show
that the stronger field moves the Alfv\'en surface, inside which the magnetic field controls the
wind, from $\sim$10--25$\,\Rsun$ to $\sim$40--60$\,\Rsun$, so CME shocks deliver over ten times
the particle fluence of flares. Because the forcing is powered by deceleration, it follows the
measured wind history ($\dot M\propto t^{-2.3}$) and declines as $\sim t^{-3}$. It was
$\sim$50--100 times stronger at 3.9~Ga than at 2~Ga and several hundred times today's level. This
decline, strongest when the Sun was faintest, is what greenhouse solutions of the faint young Sun
(FYS) problem usually impose by hand. The magnitude is small: SEP-driven $\mathrm{N_2O}$ adds at
most a few $\unit{W\,m^{-2}}$ to a 20--40$\unit{W\,m^{-2}}$ greenhouse requirement, and only if a
fractal organic haze extends its lifetime. Counting energy deposited below a survival level near
1~hPa, we find the particles fix $\sim$$4\e{10}$--$4\e{11}\unit{mol\,N\,yr^{-1}}$, which could
supply prebiotic feedstock to surface pools. The largest uncertainties are $E_b$, the event rate,
the survival depth, and the $\mathrm{N_2O}$ lifetime.
\end{abstract}


\section{Introduction}
\label{sec:intro}

Standard stellar-evolution models require the Sun to have been $\sim$25--30\% fainter during the
Archean than today. Yet geological evidence for liquid water and early life points to a clement,
largely ice-free early Earth \citep{sagan1972,feulner2012}. Early Mars poses an even sharper
version of the problem: it shows evidence of liquid surface water while receiving less than half
of Earth's insolation \citep{feulner2012}. No consensus has yet been reached on how this ``faint
young Sun'' (FYS) paradox is resolved \citep{charnay2020,catling2020}. The lower luminosity
corresponds to a radiative forcing deficit of $\approx60\unit{W\,m^{-2}}$ at 3.8~Ga
\citep{feulner2012}. Plausible changes in cloud cover and surface albedo can offset part of this
deficit, but not more than $\sim$60\% of it \citep{charnay2020}, which leaves of order
$20$--$40\unit{W\,m^{-2}}$ for an enhanced greenhouse effect to supply. A viable resolution must
also let that forcing fade as the solar luminosity rose; otherwise the planet would overheat
later in the Archean. Proposed solutions invoke either a stronger greenhouse effect, from
CO$_2$, CH$_4$, N$_2$O, or reduced-gas hazes \citep{haqqmisra2008,wolftoon2013,wordsworth2013},
or a lower planetary albedo. Each must satisfy independent geochemical constraints, and few
studies address how the forcing declines with time.

The young Sun also differed from the present Sun in its magnetic activity. It rotated several
times faster, and its stronger dynamo drove a denser, faster wind and far more frequent and
energetic flares and coronal mass ejections (CMEs) \citep{skumanich1972,ribas2005,wood2005,gudel2007}.
Superflare statistics from Kepler and TESS, together with direct detections of stellar CMEs on
young solar analogues, imply eruption and particle-event rates orders of magnitude above present
\citep{namekata2022,namekata2024}. For the large, fast events that matter here we adopt
$\sim$30--300 per year (\S\ref{sec:sepprod}). The heightened activity also changes where the wind
comes from. On the present Sun near activity minimum, the fastest wind flows from large polar
coronal holes, where the magnetic field opens into interplanetary space. In a persistently active
corona these polar holes are suppressed, and the outflow is dominated by the Alfv\'enic slow
wind. This is a slow wind that carries the large-amplitude Alfv\'enic fluctuations usually seen in
fast wind. It originates at coronal-hole boundaries and in pseudostreamers, coronal structures
that separate open-field regions of the same magnetic polarity \citep{Panasenco2020}. We show in
\S\ref{sec:alfven} that this wind topology also affects the size of the Alfv\'en surface, the
boundary inside which the Sun's magnetic field controls the wind. Magnetic activity of this kind
is now recognized as a major driver of early atmospheric chemistry and of the prebiotic
environment \citep{airapetian2016,airapetian2020}.

Solar energetic particles (SEPs) carry much of that influence. SEPs that penetrate into the
middle atmosphere dissociate N$_2$ and O$_2$. This drives the fixation of nitrogen into NO$_x$ and
eventually into N$_2$O, a potent greenhouse gas. The same process generates HCN and related
feedstocks for prebiotic synthesis \citep{airapetian2016,sur2025}. SEPs come in two physically
distinct classes \citep{reames2013,desai2016}. \emph{Impulsive} events are accelerated by
reconnection in flares and jets. \emph{Gradual} events are accelerated by CME-driven shocks and
produce the largest fluences near 1~AU. Nobody has yet quantified which class dominates the
young-Sun forcing, or how its particles reach Earth's atmosphere through a very different
heliosphere and magnetosphere.

The Alfv\'en surface strongly affects that delivery. Inside this surface the magnetic field
dominates the wind and can confine eruptions and energetic particles. Parker Solar Probe (PSP)
now crosses the surface directly and places it, rugged and variable in time, at
$\sim$10--25$\,\Rsun$ for the present Sun \citep{kasper2021, liu2023, finley2025}. Around the
young Sun, the much stronger field pushed the surface far outward. The resulting larger magnetic
``cage'' changed how CMEs escape and how SEPs travel to Earth.

The goal of this paper is to quantify the energetic-particle forcing that young-Sun CMEs
delivered to the early Earth and to follow how that forcing changed as the Sun aged. We then ask
what it implies for the FYS problem and for the supply of prebiotic nitrogen. To do this we build
a single, semi-analytic model chain (\S\ref{sec:approach}). A fast CME
($v_0\sim3\text{--}4\e{3}\unit{km\,s^{-1}}$) launched from the young Sun decelerates against the
dense inner heliosphere and dissipates $\Delta E_k\sim6\e{25}\unit{J}$. Diffusive shock
acceleration (DSA) converts a fraction of this energy into SEPs, whose spectral hardness, captured
by a broken-power-law break energy $E_b$, sets how deep they reach into the atmosphere. We follow
the particles through the enlarged Alfv\'en cage and the tightly wound young-Sun Parker spiral and
compare the impulsive and gradual channels. We then compute the atmospheric forcing that results.

We find that the time-dependence of this forcing is its most secure property. Because the forcing
is powered by deceleration, it inherits the measured wind history, $\dot M\propto t^{-2.3}$, and
declines as $\sim t^{-3}$. It was therefore $\sim$50--100 times stronger at 3.9~Ga than at 2~Ga,
and several hundred times stronger than today (\S\ref{sec:thermostat}). The solar luminosity rises
for an independent reason, core hydrogen burning. Both trends are monotonic in stellar age, so the
forcing was strongest when the Sun was faintest and weakened as the Sun brightened. Greenhouse
resolutions of the FYS problem need such a decline but usually impose it by hand. The decline
follows from the wind history alone, so it holds across the full parameter uncertainty and can be
tested against the rock record.

The magnitude of the forcing is small. Independent radiative--climate modelling finds that
greenhouse species produced by space weather warm the surface by $\lesssim0.3\unit{K}$ in steady
state, and that $\mathrm{N_2O}$ contributes little of this \citep{biasiotti2026}. Our forcing is
transient and depends on haze, a regime that steady-state, haze-free models do not cover. In that
regime SEP-driven $\mathrm{N_2O}$ adds to the greenhouse budget without closing it. We evaluate its
net warming explicitly, including the cooling by the haze it requires. The same particle flux
supplies prebiotic feedstock, so the climate and origin-of-life consequences share one driver. All
results are order-of-magnitude estimates. The uncertainty budget (Table~\ref{tab:robust}) shows
that most of the uncertainty comes from $E_b$, the event frequency, the $\mathrm{N_2O}$ lifetime,
and the depth below which fixed nitrogen survives, and it lists the measurements and simulations
that would reduce it.

Section~\ref{sec:approach} describes the model chain and how each figure is produced.
Section~\ref{sec:environment} sets out the young-Sun environment, and Section~\ref{sec:cme} the
CME deceleration energetics and SEP production. Section~\ref{sec:alfven} treats the Alfv\'en
surface and its consequences for CME caging and particle escape, and Section~\ref{sec:transport}
the SEP transport and decay. Sections~\ref{sec:fys} and \ref{sec:ool} discuss the FYS and
origin-of-life implications. Section~\ref{sec:synthesis} gives a synthesis with testable
predictions, and Section~\ref{sec:conclusions} concludes.

\section{Approach: a semi-analytic model chain}
\label{sec:approach}

The results in this paper come from a chain of linked calculations, each simple enough to check
by hand or with a short script. Most links are closed-form physical relations, such as energy
conservation for the decelerating CME and the Rankine--Hugoniot jump condition at its shock.
Others are one-dimensional numerical integrations of standard equations, for example the
drag-based equation of motion and the field-aligned diffusion of energetic particles. The
atmospheric response uses order-of-magnitude scalings and a simple band model for the radiative
forcing. The whole chain runs in seconds, so we can scan the uncertain parameters and build the
uncertainty budget of Table~\ref{tab:robust}.

The chain has four modules, each passing a small set of numbers to the next. \emph{Module~1}
(\S\ref{sec:decel}) integrates the deceleration of the CME and evaluates the strength of the shock
it drives. It returns the dissipated energy $\Delta E_k$, the shock compression history, and the
break energy $E_b$. \emph{Module~2} (\S\ref{sec:sepprod}) converts a fraction of $\Delta E_k$ into
a broken-power-law SEP spectrum spread over the shock-connected solid angle. It then computes the
energy intercepted by Earth and the fraction deposited below a survival pressure $p_s$, the level
below which fixed nitrogen survives. \emph{Module~3} (\S\S\ref{sec:alfven}--\ref{sec:transport})
computes the Alfv\'en surface and the transport of the two SEP classes to 1~AU, which gives the
delivered fluence. \emph{Module~4} (\S\S\ref{sec:fys}--\ref{sec:ool}) converts the deposited
energy into fixed nitrogen and evaluates the radiative forcing and the prebiotic feedstock
delivered to the surface, both as functions of time. Each module could be replaced by a community
code: a global heliospheric MHD model such as EUHFORIA for Module~1, a focused-transport code such
as iPATH or EPREM for Module~3, and a coupled photochemical--climate model for Module~4.

\begin{figure*}[t]
\centering
\includegraphics[width=\textwidth]{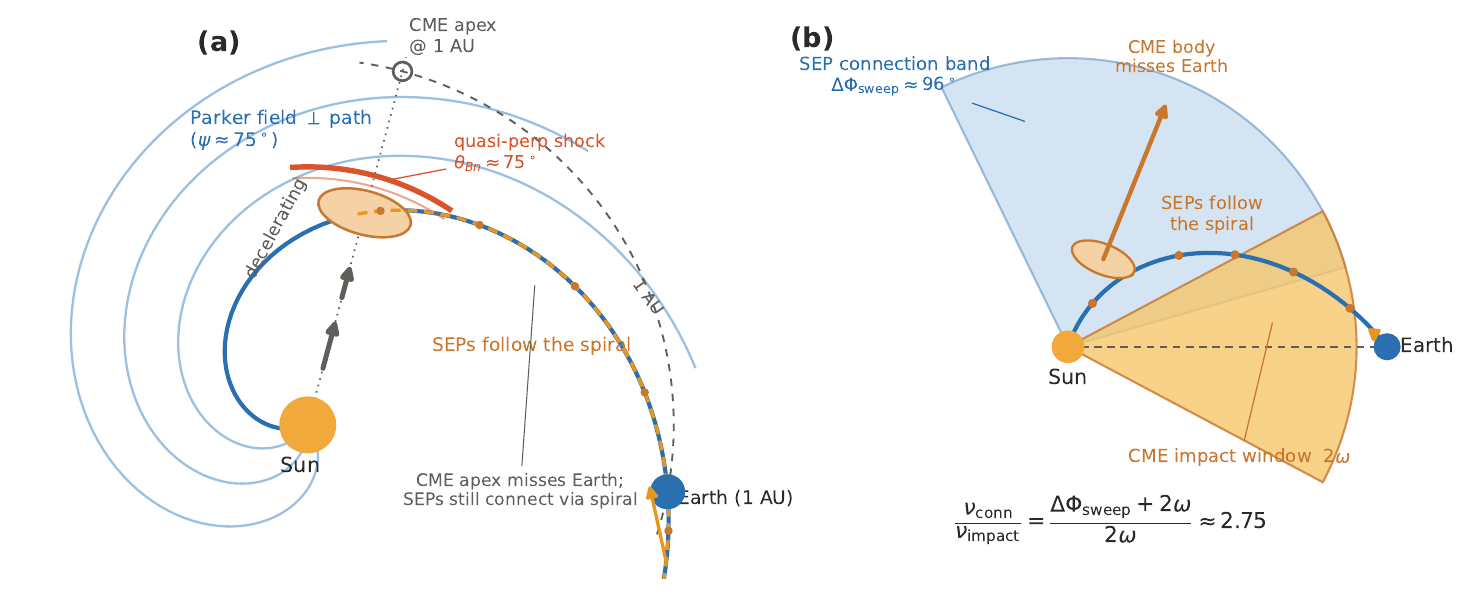}
\caption{Schematic of the mechanism. \textbf{(a)}~Global view. A CME launched from the young
Sun propagates radially and decelerates in the dense inner heliosphere (Fig.~\ref{fig:decel}),
driving a shock (coral) that accelerates SEPs. Because the young-Sun Parker spiral is tightly
wound ($\psi\approx75^\circ$; Eq.~\ref{eq:psi}), the ambient field crosses the radial CME path
almost perpendicularly, giving a quasi-perpendicular shock ($\theta_{Bn}\approx75^\circ$). The
SEPs follow the spiral to Earth, which sits $\sim85^\circ$ in longitude from the CME apex, so
the ejecta can miss Earth while its particles still connect. \textbf{(b)}~Connection versus
impact (\S\ref{sec:decouple}). In source longitude the CME impact window (amber, width
$2\omega$) is narrow, because the apex must point within $\omega$ of the Sun--Earth line. The
connection band (blue, $\Delta\Phi_{\rm sweep}\approx96^\circ$), swept out as the shock advances
through its acceleration zone, is far broader. A CME whose body propagates radially and
misses Earth still irradiates it through its spiral-connected shock; the forcing-relevant
occurrence rate is therefore the connection rate, larger than the impact rate by
$\nu_{\rm conn}/\nu_{\rm impact}\approx2.75$ (Eq.~\ref{eq:ratio}).}
\label{fig:cartoon}
\end{figure*}

Table~\ref{tab:figs} gives the method and inputs behind each figure. Apart from the schematic of
Fig.~\ref{fig:cartoon}, every figure is computed directly from the equations and parameter values
in the text. All calculations use Python with NumPy, and the figures are drawn with Matplotlib.
The scripts that implement the four modules and generate Figs.~\ref{fig:decel}--\ref{fig:regime}
will be made publicly available on publication.

\begin{table}[ht]
\centering
\caption{How each figure is produced. ``Analytic'' curves evaluate closed-form expressions;
``numerical'' curves integrate the listed equations on a grid.}
\label{tab:figs}
\small
\begin{tabular}{@{}llll@{}}
\toprule
\cell{0.06}{Fig.} & \cell{0.20}{Content} & \cell{0.34}{Method and equations} & \cell{0.30}{Main inputs} \\
\midrule
\cell{0.06}{\ref{fig:cartoon}} & \cell{0.20}{Mechanism and connection geometry} &
\cell{0.34}{Schematic; angles from Eqs.~\eqref{eq:psi} and \eqref{eq:ratio}} &
\cell{0.30}{$\psi=75^\circ$, $\Delta\Phi_{\rm sweep}\approx96^\circ$} \\[2pt]
\cell{0.06}{\ref{fig:decel}} & \cell{0.20}{CME speed and shock compression vs.\ distance} &
\cell{0.34}{Numerical: drag-based equation of motion, Eq.~\eqref{eq:dbm}, from 0.1 to 1~AU;
Rankine--Hugoniot compression along the path (Module~1)} &
\cell{0.30}{$M=10^{13}\unit{kg}$, $v_0=3500$, $w=600\unit{km\,s^{-1}}$,
$\Gamma=10^{-7}\unit{km^{-1}}$; $v_{\rm fms}$ profile and $\mathcal{M}_{\rm crit}=1.3$ as in
\S\ref{sec:decel}} \\[2pt]
\cell{0.06}{\ref{fig:spectrum}} & \cell{0.20}{SEP energy spectrum} &
\cell{0.34}{Analytic: broken power law, Eq.~\eqref{eq:broken}; top axis: pressure level reached,
from the proton range--energy relation $R=2.3\e{-3}\,E^{1.82}\unit{g\,cm^{-2}}$ ($E$ in MeV);
band: survival threshold $E_s$ for $p_s=0.15$--3~hPa} &
\cell{0.30}{$\gamma_1=1.5$, $\gamma_2=4$, $E_b=30\unit{MeV}$} \\[2pt]
\cell{0.06}{\ref{fig:fpen}} & \cell{0.20}{Energy deposited below $p_s$ vs.\ $E_b$} &
\cell{0.34}{Numerical quadrature of Eq.~\eqref{eq:fpen} (Module~2)} &
\cell{0.30}{$\gamma_1=1.5$, $\gamma_2=3$--5, $p_s=0.15$, 1, 3~hPa, $E_{\min}=0.1\unit{MeV}$,
$E_{\max}=1\unit{GeV}$} \\[2pt]
\cell{0.06}{\ref{fig:alfven}} & \cell{0.20}{Alfv\'en surface, modern vs.\ 3.9~Ga} &
\cell{0.34}{Numerical: radial profiles from Eqs.~\eqref{eq:vA}--\eqref{eq:vwind}; Alfv\'en point
from $v=v_A$ (Module~3)} &
\cell{0.30}{($v_\infty$, $B_0$, $\dot M/\dot M_\odot$) $=$ (400$\unit{km\,s^{-1}}$, 1.4~G, 1)
and (650$\unit{km\,s^{-1}}$, 55~G, 80); acceleration scale $h_v=10$ and $13\,\Rsun$; band from $\dot M/\dot M_\odot=50$--120} \\[2pt]
\cell{0.06}{\ref{fig:sepdecay}} & \cell{0.20}{SEP intensity and fluence at 1~AU} &
\cell{0.34}{Numerical: diffusion Green's function, Eq.~\eqref{eq:green}, and its convolution
over a moving shock source, Eqs.~\eqref{eq:Iflare}--\eqref{eq:Icme} (Module~3)} &
\cell{0.30}{Table~\ref{tab:transport}; $\sim$10~MeV protons} \\[2pt]
\cell{0.06}{\ref{fig:netforcing}} & \cell{0.20}{Net $\mathrm{N_2O}$ and haze forcing} &
\cell{0.34}{Band model of \S\ref{sec:netforcing} (Module~4)} &
\cell{0.30}{$M_0=4$~ppb, $c_N=0.3\unit{W\,m^{-2}\,ppb^{-1/2}}$, $M_{\rm bg}=0.3$~ppb,
$k_h=15\unit{W\,m^{-2}}$, $R=15$ (fractal) or 2.5 (spherical)} \\[2pt]
\cell{0.06}{\ref{fig:thermostat}} & \cell{0.20}{Forcing and deficit vs.\ time} &
\cell{0.34}{Analytic: deficit $\propto 1-L(t)/L_\odot$ from the standard luminosity history
\citep[Eq.~1 of][]{feulner2012}; SEP forcing $\propto t^{-3}$ (\S\ref{sec:thermostat}); both
normalized at 3.9~Ga} &
\cell{0.30}{Solar age 4.57~Gyr; exponent $3\pm0.5$} \\[2pt]
\cell{0.06}{\ref{fig:regime}} & \cell{0.20}{Regime map in ($E_b$, $\nu$)} &
\cell{0.34}{On a grid: $\dot N_{\rm fix}=3.3\e{9}\,\fdep(p_s;E_b)\,\nu$ from
Eqs.~\eqref{eq:fpen} and \eqref{eq:globalN} (Module~4)} &
\cell{0.30}{$p_s=1$~hPa (dashed: 0.15 and 3~hPa); thresholds $10^{10}$ (pools reach $\mu$M) and
$10^{12}\unit{mol\,N\,yr^{-1}}$ (FYS closure)} \\
\bottomrule
\end{tabular}
\end{table}

\section{The young-Sun environment}
\label{sec:environment}

The plasma environment of the young Sun is reconstructed primarily from observations of
solar analogues. Stellar rotation, and with it magnetic activity, decays through
angular-momentum loss \citep{skumanich1972}, and the ``Sun in Time'' program established that
the early Sun was a rapid rotator with X-ray and extreme-UV output up to one to two orders of
magnitude above present \citep{ribas2005,gudel2007}. Astrospheric Ly$\alpha$ absorption
measurements yield a mass-loss history that declines steeply with age,
$\dot M \propto t^{-2.3}$ \citep{wood2002,wood2005}, implying a denser and faster young wind and
a correspondingly tighter Parker spiral \citep{parker1958}. We adopt these constraints as the
ambient boundary conditions for the eruptions and energetic particles treated in
\S\ref{sec:cme}.

\subsection{Young Sun boundary conditions at 3.9~Ga}
\label{sec:bc}

At a stellar age of $\sim 0.6\unit{Gyr}$ the Sun differs from today in four ways that all
bear on space weather:
\begin{enumerate}
  \item \textit{Luminosity:} $L_\odot(3.9\,\mathrm{Ga}) \approx 0.70\text{--}0.75\,L_{\odot,0}$
  (standard solar models). This is the origin of the FYS deficit (\S\ref{sec:intro}).
  \item \textit{Rotation:} young solar analogues rotate $\sim 3\times$ faster, with periods
  $P \approx 5\text{--}9\unit{d}$, i.e. $\Omega \approx (0.8\text{--}1.5)\e{-5}\unit{rad\,s^{-1}}$.
  \item \textit{Wind / mass loss:} astrospheric measurements give
  $\dot{M}\propto t^{-2.3}$ \citep{wood2005}, so the inner-heliosphere density at fixed $r$ scales
  roughly as $n_{\mathrm{amb}}\propto t^{-2.3}$; the young wind was both faster
  ($\vsw \sim 500\text{--}700\unit{km\,s^{-1}}$ baseline) and far denser.
  \item \textit{Surface field:} $\sim 10\text{--}100\unit{G}$ mean field pushes the Alfv\'en
  surface to $40\text{--}60\,\Rsun$ and tightens the Parker spiral (below).
\end{enumerate}

The adopted baseline ($v_\infty\sim600\text{--}700\unit{km\,s^{-1}}$) follows from magnetic
topology. Some studies assume an extremely fast wind for active stars, but the very fast
($\sim$750--800$\unit{km\,s^{-1}}$) wind requires the large, stable polar coronal holes of low
activity. In a persistently active corona these holes are suppressed, and the Alfv\'enic slow wind
from pseudostreamers, coronal-hole boundaries, and the separatrix web dominates the outflow
\citep{PanasencoVelli2010,PanasencoVelli2013,Panasenco2019,Panasenco2020}. A fast-rotating young
Sun is effectively always in a high-activity state, so it should drive a moderately fast,
topology-controlled wind. This supports our baseline of $\sim$600$\unit{km\,s^{-1}}$.

\subsection{The range of Parker spiral angles at 1~AU}
\label{sec:spiral}

The spiral winding (garden-hose) angle at heliocentric distance $r$ measured from the radial
direction is
\begin{equation}
  \tan\psi \;=\; \frac{\Omega\, r}{\vsw}.
  \label{eq:psi}
\end{equation}
Because both $\Omega$ ($P=5\text{--}9\unit{d}$) and $\vsw$ ($500\text{--}700\unit{km\,s^{-1}}$)
are uncertain, $\psi$ at $r=1\unit{AU}=1.5\e{11}\unit{m}$ spans the box in
Table~\ref{tab:psi}: $\psi \approx 60^\circ\text{--}77^\circ$, central value $\sim 70^\circ$
(versus $45^\circ$ today).

\begin{table}[h]
\centering
\caption{Spiral angle $\psi$ at 1~AU from Eq.~\eqref{eq:psi} for the young Sun.}
\label{tab:psi}
\begin{tabular}{lccc}
\toprule
 & $\vsw=500$ & $\vsw=600$ & $\vsw=700\unit{km\,s^{-1}}$ \\
\midrule
$P=5\unit{d}$ & $77.1^\circ$ & $74.6^\circ$ & $72.2^\circ$ \\
$P=7\unit{d}$ & $72.2^\circ$ & $68.9^\circ$ & $65.8^\circ$ \\
$P=9\unit{d}$ & $67.6^\circ$ & $63.7^\circ$ & $60.0^\circ$ \\
\bottomrule
\end{tabular}
\end{table}

Two geometric quantities follow. The magnetically well-connected source longitude is offset
westward by
\begin{equation}
  \Delta\phi \;=\; \frac{\Omega r}{\vsw} \;=\; \tan\psi \quad[\mathrm{rad}],
\end{equation}
and the field-line (Archimedean) path length from Sun to Earth is
\begin{equation}
  L \;=\; \frac{r}{2}\left[\sqrt{1+s^2} + \frac{\sinh^{-1}s}{s}\right],
  \qquad s \equiv \tan\psi = r/a,\quad a=\vsw/\Omega .
  \label{eq:path}
\end{equation}
Evaluated at representative angles (Table~\ref{tab:geom}), the connection point wraps from
$\sim 123^\circ$ to beyond $180^\circ$ west, and the path lengthens from $1.5$ to $2.2\unit{AU}$.
The longer, more tightly wound tube costs $\lesssim 2\times$ in peak SEP intensity
through adiabatic deceleration and scattering, but little in time-integrated fluence.

\begin{table}[h]
\centering
\caption{Connection longitude and field-line path length vs.\ spiral angle.}
\label{tab:geom}
\begin{tabular}{lccc}
\toprule
$\psi$ & $\tan\psi$ & $\Delta\phi$ & $L$ (AU) \\
\midrule
$65^\circ$ & $2.145$ & $122.9^\circ$ & $1.53$ \\
$70^\circ$ & $2.747$ & $157.4^\circ$ & $1.78$ \\
$75^\circ$ & $3.732$ & $213.8^\circ$ & $2.20$ \\
\bottomrule
\end{tabular}
\end{table}

\section{CME deceleration and SEP production}
\label{sec:cme}

Direct evidence for young-Sun eruptive activity has only recently emerged:
superflare-associated filament and prominence eruptions on the solar analogue EK~Draconis
\citep{namekata2022,namekata2024}, and the first multi-temperature stellar CME signature,
confirming Carrington-class and larger events \citep{namekata2025}. Whether such ejecta escape
the strong young-Sun field is debated, since an overlying large-scale field can confine weaker
CMEs \citep{alvarado2018,drake2013}; the fast, massive events treated here are those that break
out. The partition of eruptive energy into flare radiation, CME kinetic energy, and energetic
particles is calibrated on modern events \citep{emslie2012}, with diffusive shock acceleration
in large gradual events producing power-law spectra whose hardness tracks shock strength
\citep{desai2016}. The magnetospheric response of the early Earth to this forcing has been
modeled for both CME \citep{airapetian2016} and corotating-interaction-region drivers
\citep{sur2025}. Building on these constraints, we power the SEP yield by CME deceleration in
the dense inner heliosphere and parametrize its atmospheric impact by the hardness of the
shock-accelerated spectrum (its break energy $E_b$).

\subsection{Base parameters and deceleration energetics}
\label{sec:decel}

\begin{figure}[h]
\centering
\includegraphics[width=0.7\textwidth]{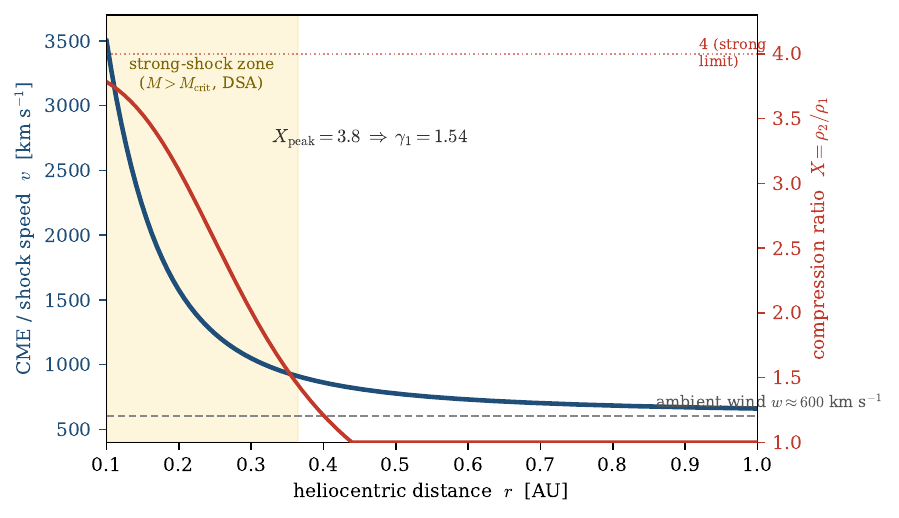}
\caption{Drag-based deceleration of the nominal young-Sun CME (blue) and the shock compression
ratio it drives (red), computed with Module~1 (\S\ref{sec:approach}; Eq.~\ref{eq:dbm}). The ejecta slows from
$v_0\approx3500$ to $\approx600\unit{km\,s^{-1}}$ across the dense inner heliosphere; the
dissipated bulk kinetic energy (Eq.~\ref{eq:dEk}) is the reservoir for particle acceleration.
The compression ratio $X\equiv\rho_2/\rho_1$ (downstream-to-upstream mass density across the
shock) is the Rankine--Hugoniot value
$X=(\gamma+1)\mathcal{M}^2/[(\gamma-1)\mathcal{M}^2+2]$ for a $\gamma=5/3$ gas, with
$\mathcal{M}$ the fast-magnetosonic Mach number; it tends to the strong-shock limit $X\to4$ at
high $\mathcal{M}$ and fixes the diffusive-shock-acceleration injection slope through
$\gamma_1=(2X+1)/[2(X-1)]\to1.5$. The shaded band is the strong-shock zone
($\mathcal{M}>\mathcal{M}_{\rm crit}$), where $X\approx X_{\rm peak}=3.8$ and acceleration is
efficient; beyond it the shock relaxes toward a wave ($X\to1$) and the source spectrum
steepens.}
\label{fig:decel}
\end{figure}

Take a large young-Sun event with mass $M \sim 10^{16}\unit{g} = 10^{13}\unit{kg}$ (a parameter;
superflare CMEs may reach $10^{17}\text{--}10^{18}\unit{g}$), launched at
$v_0 = 3.5\e{6}\unit{m\,s^{-1}}$ and decelerated against the dense inner heliosphere to the
ambient baseline $v_1 = 6\e{5}\unit{m\,s^{-1}}$. The bulk kinetic energy dissipated in transit
is
\begin{align}
  \Delta E_k &= \tfrac{1}{2}M\left(v_0^2-v_1^2\right) \nonumber\\
  &= \tfrac{1}{2}(10^{13})\big[(3.5\e{6})^2-(6\e{5})^2\big] \nonumber\\
  &= \tfrac{1}{2}(10^{13})(1.189\e{13}) \;\approx\; 6\e{25}\unit{J}.
  \label{eq:dEk}
\end{align}
This is several Carrington events' worth dumped in a single transit, and since
$v_1^2 \ll v_0^2$ almost the entire launch energy is available. The deceleration history and
the shock compression ratio it drives are shown in Fig.~\ref{fig:decel}.

The deceleration in Fig.~\ref{fig:decel} is computed with the drag-based model
\citep{vrsnak2013}, in which the ejecta is coupled aerodynamically to the ambient wind:
\begin{equation}
  \frac{dv}{dt} = -\Gamma\,(v-w)\,|v-w| , \qquad \frac{dr}{dt} = v ,
  \label{eq:dbm}
\end{equation}
where $w=600\unit{km\,s^{-1}}$ is the ambient wind speed and $\Gamma$ is the drag parameter. We
integrate Eq.~\eqref{eq:dbm} from $r=0.1$ to $1\unit{AU}$ with $v_0=3500\unit{km\,s^{-1}}$ and
$\Gamma=10^{-7}\unit{km^{-1}}$, a value within the range fitted to modern CMEs
\citep{vrsnak2013}; the denser young wind would, if anything, raise $\Gamma$. At each step the
fast-magnetosonic Mach number is $\mathcal{M}=(v-w)/v_{\rm fms}(r)$, with an assumed ambient
profile $v_{\rm fms}=400\unit{km\,s^{-1}}\,(r/0.1\,\mathrm{AU})^{-0.4}$. The compression ratio
$X(r)$ then follows from the Rankine--Hugoniot relation given in the caption of
Fig.~\ref{fig:decel}, and the strong-shock zone is where $\mathcal{M}>\mathcal{M}_{\rm crit}=1.3$.
Three points support the use of this calculation. First, the energy reservoir of
Eq.~\eqref{eq:dEk} depends only on the endpoints $v_0$ and $v_1$, so it does not depend on the
deceleration path or on the propagation model. Second, the drag-based model is validated for the
deceleration profile $v(r)$ itself. Compared with the WSA--ENLIL+Cone MHD model, it reproduces
bulk kinematics and arrival times within the MHD model's own scatter
\citep{vrsnak2014,dumbovic2018}, so a full magnetohydrodynamic integration would change neither
$v(r)$ nor $\Delta E_k$ much. Third, the compression $X(r)$ is the one model-dependent element.
The drag-based model tracks the ejecta and not its shock, so $X(r)$ in Fig.~\ref{fig:decel} is an
auxiliary Rankine--Hugoniot estimate evaluated from an assumed ambient fast-magnetosonic and field
profile. The jump condition is exact, but its upstream input is assumed, so $X(r)$ and the
injection slope $\gamma_1$ derived from it carry a model dependence that only a global MHD
solution removes. We therefore treat $v(r)$ and $\Delta E_k$ as well determined and $X(r)$ as a
calibrated estimate. A heliospheric MHD run (e.g.\ EUHFORIA) driven by the young-Sun boundary
would test the shock obliquity and compression and is a well-defined task for follow-up work.
Fast CMEs launched near $v_0\approx3500\unit{km\,s^{-1}}$ are also known to decelerate more
sharply than such models capture \citep{dumbovic2018}. That bias moves the energy deposition and
the strongest shock further inward, which if anything favours the inner-heliosphere SEP source
used here.

\paragraph{The deceleration energy as the SEP reservoir.}
The dissipated bulk kinetic energy of Eq.~\eqref{eq:dEk} is the correct reservoir for
diffusive shock acceleration and involves no double counting. The work the ejecta does against
drag is the energy transferred across its bow shock to the swept-up plasma, and DSA returns a
fraction $\varepsilon$ of it to suprathermal particles. Using $\Delta E_k$ as the budget and
$\varepsilon$ as the efficiency therefore reproduces the standard scaling of SEP energy with CME
kinetic energy \citep{emslie2012}. The deceleration framing adds a model-independent anchor (the
endpoints) and a time-dependence inherited from the wind history (\S\ref{sec:thermostat}).
Because $v_1^2\ll v_0^2$, $\Delta E_k$ is within a few percent of the launch kinetic energy and is
an upper bound on what the shock can process; all conversion losses are folded into
$\varepsilon$. The required power is also affordable. At $\nu\sim100\unit{yr^{-1}}$ the
time-averaged CME power is $\nu\,\Delta E_k\sim2\e{20}\unit{W}$, about $10^{-3}$ of the young
Sun's X-ray and EUV luminosity and $10^{-6}$ of its bolometric output.

\subsection{SEP production by DSA during propagation}
\label{sec:sepprod}

\paragraph{Step 1: energy into particles.}
A strong CME-driven shock channels a fraction $\varepsilon \sim 0.1$ into SEPs. This value is
calibrated on modern large gradual events \citep{emslie2012} (the standard 1--20\% range);
transferring it to the young Sun is an assumption, though the dense, high-Mach young heliosphere
plausibly sits at the favourable end. Thus
\begin{equation}
  \Esep = \varepsilon\,\Delta E_k \approx 0.1\times 6\e{25} \approx 6\e{24}\unit{J}.
\end{equation}

\paragraph{Step 2: beaming and the fluence at Earth.}
SEPs are confined to the shock-connected flux tubes. A wide fast shock
($\sim 120^\circ$ longitude $\times\,\pm30^\circ$ latitude) fills a solid angle
\begin{equation}
  \Omega_{\mathrm{ill}} \approx \Delta\phi\,[\sin\lambda]_{-30}^{+30}\approx 2.1\unit{sr},
  \qquad b \equiv \frac{4\pi}{\Omega_{\mathrm{ill}}}\approx 6,
\end{equation}
so the energy fluence in the connected patch is
\begin{equation}
  F_E = \frac{\Esep}{\Omega_{\mathrm{ill}}\,r^2}
  = \frac{6\e{24}}{2.1\times(1.5\e{11})^2}
  \approx 1.3\e{-2}\unit{J\,cm^{-2}} \approx 8\e{10}\unit{MeV\,cm^{-2}} .
  \label{eq:FE}
\end{equation}
The total SEP energy intercepted by Earth's atmosphere is
\begin{equation}
  E_{\mathrm{atm}} = F_E\,\pi\RE^2 = 1.3\e{-2}\times 1.29\e{18}\unit{cm^2}
  \approx 1.6\e{16}\unit{J\;per\;event}.
  \label{eq:Eatm}
\end{equation}

\paragraph{Magnetospheric access.}
Equation~\eqref{eq:Eatm} gives the energy intercepted by Earth. How much of it precipitates
depends on the early geomagnetic field, which sets the rigidity cutoff and the area of the open
polar cap through which SEPs reach the atmosphere. The 3.9~Ga field is poorly constrained.
Palaeomagnetic data place a geodynamo by $\sim$3.4--4.2~Ga but leave its strength uncertain. A
weaker or more multipolar field, compressed further by the dense young-Sun wind and frequent CME
impacts, would enlarge the open polar cap and admit particles over a larger fraction of the
globe, raising access above the present value \citep{airapetian2016,sur2025}.
\citet{biasiotti2026} reach the same conclusion from the atmospheric-chemistry side. They note
that during polarity reversals, weak-field epochs, or strongly quadrupolar configurations, SEPs
reaching mid and low latitudes could have amplified the chemical and climatic response well
beyond the present-field case they adopt. Pulling the other way, the Moon had its own dynamo at
this epoch and orbited at only $\sim$18$\,\RE$, so the coupled Earth--Moon magnetospheres may
have shielded the early Earth better than its field alone \citep{green2020}. That shielding was
demonstrated against bulk solar-wind stripping, however. Rigid ($\gtrsim$10$\unit{MeV}$) SEPs
precipitate through the polar open-field region and are much less sensitive to the magnetopause
standoff that deflects the thermal wind, so the direct effect on the SEP budget is limited. The
net access factor is therefore uncertain in both directions. We absorb it into a multiplier
$f_{\rm acc}\sim1$ on $E_{\mathrm{atm}}$, set to unity. A proper treatment would couple the SEP
flux to a co-evolving palaeo-magnetospheric model that includes the lunar dynamo, which peaked
near 4~Ga and faded by $\sim$1.5~Ga.

\paragraph{Step 3: broken-power-law spectrum and deposition below the survival level.}
A single power law with a hard cutoff at $E_{\max}$ is unphysical: the maximum energy is set by
the finite time the shock remains strong, which produces a gradual rollover in the spectrum. We
therefore adopt a broken (double) power law,
\begin{equation}
  \frac{d\Phi}{dE}=K
  \begin{cases}
    E^{-\gamma_1}, & E_{\min}\le E\le E_b,\\[1ex]
    E_b^{\,\gamma_2-\gamma_1}\,E^{-\gamma_2}, & E_b< E\le E_{\max},
  \end{cases}
  \label{eq:broken}
\end{equation}
continuous at the break $E_b$, with a hard low-energy slope $\gamma_1\approx1.5$ (strong-shock
DSA, $f(p)\propto p^{-4}\Rightarrow dN/dE\propto E^{-1.5}$) and a steep rollover
$\gamma_2\approx3$--$5$. The break energy itself follows from the acceleration time:
$t_{\rm acc}(E_b)=t_{\rm strong}$, with $t_{\rm acc}\propto\kappa(E)/u_{\rm sh}^2$ (Bohm
diffusion $\kappa\propto E/B$), so
\begin{equation}
  E_b \;\propto\; u_{\rm sh}^2\,B\,t_{\rm strong}.
  \label{eq:Eb}
\end{equation}
The young Sun's fast shocks and strong inner-heliosphere field both push $E_b$ up (tens to
hundreds of MeV), favouring penetration; $E_b$ follows directly from the Module~1 Mach-number
history $X(r)$ of Fig.~\ref{fig:decel}, to be refined by a global MHD integration
(\S\ref{sec:decouple} and Fig.~\ref{fig:decel} caption).

Not all of the intercepted energy matters chemically. Odd nitrogen produced high in the
atmosphere is destroyed there: NO is photolysed, and the released N atoms remove further NO
through $\mathrm{N+NO\to N_2+O}$. Products formed deeper can survive, be mixed downward, and rain
out. We therefore introduce a survival pressure $p_s$, the level below which fixed nitrogen is
taken to survive, and compute the fraction of SEP energy deposited at pressures greater than
$p_s$. We use the proton range--energy relation $R(E)=a_R E^{\beta_R}$, with
$a_R=2.3\e{-3}\unit{g\,cm^{-2}}$ and $\beta_R=1.82$ ($E$ in MeV), appropriate for an
$\mathrm{N_2}$-dominated atmosphere. A proton arriving vertically reaches the column
$N_s=p_s/g$ only if its energy exceeds $E_s=(N_s/a_R)^{1/\beta_R}$; it then crosses that level
with residual energy $(E^{\beta_R}-E_s^{\beta_R})^{1/\beta_R}$, all of which is deposited below
it. The deposited fraction is
\begin{equation}
  \fdep(p_s;E_b)=
  \frac{\int_{E_s}^{E_{\max}}\big(E^{\beta_R}-E_s^{\beta_R}\big)^{1/\beta_R}\,(d\Phi/dE)\,dE}
       {\int_{E_{\min}}^{E_{\max}}E\,(d\Phi/dE)\,dE},
  \label{eq:fpen}
\end{equation}
with $d\Phi/dE$ from Eq.~\eqref{eq:broken}. For $p_s=0.15$, 1, and 3~hPa the threshold energy is
$E_s\approx10$, 28, and 52~MeV; in today's atmosphere these levels lie near 62, 48, and 40~km.
The survival pressure is set by photochemistry that we do not model, so we treat it as a
parameter in the range $p_s=0.15$--3~hPa, with a nominal 1~hPa, close to the present
stratopause. We use pressure as the vertical coordinate because an anoxic Archean atmosphere had no
ozone-heated stratosphere, so its temperature profile, and hence the altitude of a given pressure
level, differed from today's. Oblique arrival at an angle $\theta$ to the vertical is equivalent
to replacing $p_s$ by $p_s/\cos\theta$ and is absorbed in this range.

\begin{figure}[h]
\centering
\includegraphics[width=0.7\textwidth]{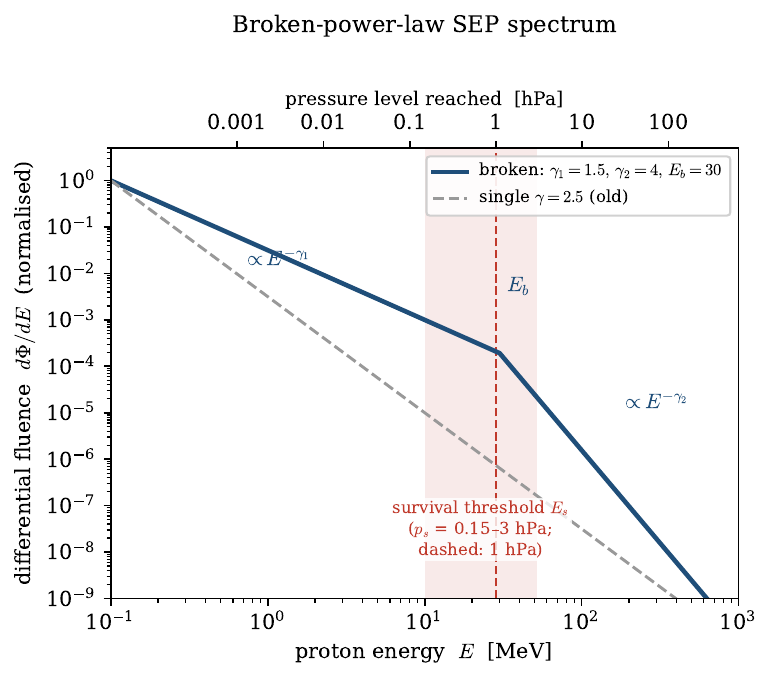}
\caption{Broken-power-law SEP spectrum (Eq.~\ref{eq:broken}; $\gamma_1=1.5$, $\gamma_2=4$,
$E_b=30\unit{MeV}$) compared with the discarded single power law. The shaded band marks the
survival threshold $E_s$ for $p_s=0.15$--3~hPa (Eq.~\ref{eq:fpen}), with the nominal
$p_s=1$~hPa dashed; only protons above $E_s$ deposit energy below the survival level. The top
axis gives the pressure level that a vertically arriving proton of energy $E$ reaches, from the
range--energy relation of \S\ref{sec:sepprod}.}
\label{fig:spectrum}
\end{figure}

\begin{figure}[h]
\centering
\includegraphics[width=0.7\textwidth]{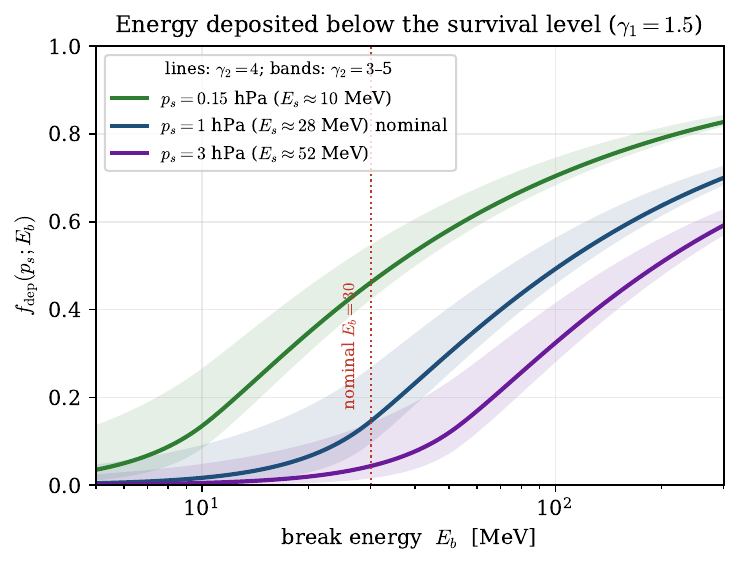}
\caption{Fraction of SEP energy deposited below the survival level, $\fdep(p_s;E_b)$
(Eq.~\ref{eq:fpen}), as a function of break energy for $p_s=0.15$, 1 (nominal), and 3~hPa. Lines
are for $\gamma_2=4$; shaded bands span $\gamma_2=3$--5. The dominant control is $E_b$ relative to
the survival threshold $E_s$; the rollover slope matters most when $E_s$ lies near the break.}
\label{fig:fpen}
\end{figure}

Two consequences follow. First, because $\gamma_2$ is steep, the high-energy segment converges
and the result depends only weakly on $E_{\max}$: changing $E_{\max}$ from 1 to 0.3~GeV lowers
$\fdep$ by at most 4\%, which removes the pathology of the single-power-law treatment. Second, the
master parameter is the break energy $E_b$ relative to the survival threshold $E_s$
(Fig.~\ref{fig:spectrum}, Table~\ref{tab:fpen}, and Fig.~\ref{fig:fpen}). For the nominal
$p_s=1$~hPa, $\fdep$ rises from 2\% at $E_b=10$~MeV to 15\% at 30~MeV and 49\% at 100~MeV.
Because $E_s$ then lies close to the break, $\fdep$ is also sensitive to the rollover slope: at
$E_b=30$~MeV it ranges from 0.10 to 0.27 for $\gamma_2=5$ to 3.

\begin{table}[h]
\centering
\caption{Energy deposited below the survival level and nitrogen yield vs.\ break energy $E_b$
($\gamma_1=1.5$, $\gamma_2=4$). $\fdep$ from Eq.~\eqref{eq:fpen} for three survival pressures;
per-event N from $N_{\mathrm{fix}}=G\,\fdep\,E_{\mathrm{atm}}$ at the nominal $p_s=1$~hPa, with
$E_{\mathrm{atm}}=1.6\e{16}\unit{J}$ and $G=1.2\e{17}$ fixed N atoms $\mathrm{J^{-1}}$.
$\Phi(>10\,\mathrm{MeV})$ is the event-integrated fluence above 10~MeV, a standard measure of
event size.}
\label{tab:fpen}
\begin{tabular}{lccccc}
\toprule
 & \multicolumn{3}{c}{$\fdep(p_s)$} & & \\
\cmidrule(lr){2-4}
$E_b$ [MeV] & 0.15~hPa & 1~hPa & 3~hPa & $\Phi(>10\,\mathrm{MeV})$ [cm$^{-2}$] & $N_{\mathrm{fix}}$ [mol/event] \\
\midrule
$10$  & $0.14$ & $0.017$ & $0.005$ & $1.2\e{9}$ & $5.5\e{7}$ \\
$30$  & $0.46$ & $0.15$  & $0.043$ & $2.0\e{9}$ & $4.7\e{8}$ \\
$100$ & $0.70$ & $0.49$  & $0.32$  & $1.5\e{9}$ & $1.6\e{9}$ \\
\bottomrule
\end{tabular}
\end{table}

\paragraph{Step 4: deep energy deposition and the global rate.}
The energy deposited below the survival level is $E_{\mathrm{dep}}=\fdep\,E_{\mathrm{atm}}$; for
the nominal $p_s=1$~hPa this is $\sim3\e{14}\unit{J}$ ($E_b=10$~MeV) to $8\e{15}\unit{J}$
($E_b=100$~MeV). To convert deposited energy into fixed nitrogen we use the standard yield
for energetic-particle ionization of $\mathrm{N_2}$-rich air: $\sim$1.25 N atoms per $\sim$35~eV
ion pair \citep{porter1976}, of which the $\sim$55\% produced in the excited N($^2$D) state set the
net production of odd nitrogen \citep{jackman2005}. This gives $\sim$0.7 fixed N atoms per ion
pair, or $G\approx1.2\e{17}\unit{J^{-1}}$. The yield is calibrated for present-day air; the
excited-state fraction is itself uncertain, and in an anoxic $\mathrm{N_2}$--$\mathrm{CO_2}$
atmosphere the conversion of excited N atoms into fixed nitrogen may differ, so we carry a factor
of $\sim$2 in $G$ (Table~\ref{tab:robust}). The per-event yield is
$N_{\mathrm{fix}}=G\,\fdep\,E_{\mathrm{atm}}$; for the nominal $E_b=30\unit{MeV}$ and
$p_s=1$~hPa, $E_{\mathrm{dep}}\approx2.3\e{15}\unit{J}$ and
$N_{\mathrm{fix}}\approx4.7\e{8}\unit{mol\,N}$. With event frequency $30\text{--}300\unit{yr^{-1}}$
for a fast-rotating young Sun, the global rate of SEP nitrogen fixation that survives below $p_s$
is
\begin{equation}
  \dot N_{\mathrm{fix}} \sim (1.4\e{10}\;\text{--}\;1.4\e{11})\unit{mol\,N\,yr^{-1}},
  \label{eq:globalN}
\end{equation}
uncertain by a factor of $\sim$3 either way across $p_s=0.15$--3~hPa. This is $\sim$3--30\% of
the modern lightning source ($\sim 5\e{11}\unit{mol\,N\,yr^{-1}}$), and at the connection rate
(Eq.~\ref{eq:globalNconn}) its upper end approaches the modern value. Because lightning was itself
weaker on the early Earth, SEP fixation was plausibly comparable to the contemporaneous lightning
source.

\subsection{SEP forcing decouples from CME impact geometry}
\label{sec:decouple}

The frequency $\nu$ in Eq.~\eqref{eq:globalN}, and every estimate that follows from it,
differs from the rate at which CMEs strike Earth. The bulk consequences of an impact, such as
magnetospheric compression and the ram pressure of the sheath and ejecta, require a geometric hit.
The atmospheric chemistry, however, is driven by the SEP flux, and energetic particles follow the
shock-connected flux tubes. A CME whose body misses Earth still irradiates it if part of its shock
is magnetically connected back along the spiral. The relevant $\nu$ is therefore a connection
rate, which is larger than the impact rate.

The two cross-sections, in ecliptic longitude, are unequal. A CME of angular half-width
$\omega$ and apex longitude $\phi_a$ (measured from the Sun--Earth line) strikes Earth if
$|\phi_a|<\omega$: an impact window $2\omega$ wide. Magnetic connection is set instead by the
Earth-rooted field line, whose footpoint at heliocentric distance $r$ lies a longitude
\begin{equation}
  \phi_c(r)=\frac{\Omega}{\vsw}\,(1\,\mathrm{AU}-r)=\tan\psi\,(1-r/1\,\mathrm{AU})\quad[\mathrm{rad}]
\end{equation}
west of the sub-Earth meridian (cf.\ Eq.~\ref{eq:psi}; $\phi_c\!\to\!\tan\psi=213.8^\circ$ as
$r\!\to\!0$, vanishing at $r=1\,$AU). As the shock advances through its effective acceleration
zone $r\in[r_0,r_a]$, $\phi_c$ sweeps a band
\begin{equation}
  \Delta\Phi_{\rm sweep}=\tan\psi\,\frac{r_a-r_0}{1\,\mathrm{AU}},
\end{equation}
and connection is realised at some instant of the transit whenever the apex falls within that
band (broadened by $\omega$). Hence
\begin{equation}
  \frac{\nu_{\rm conn}}{\nu_{\rm impact}}=\frac{\Delta\Phi_{\rm sweep}+2\omega}{2\omega}
  =1+\frac{\tan\psi}{2\omega}\,\frac{r_a-r_0}{1\,\mathrm{AU}}.
  \label{eq:ratio}
\end{equation}
For $\psi=75^\circ$ ($\tan\psi=3.73$) and a strong-shock zone $r_0\simeq0.05$ to
$r_a\simeq0.5\unit{AU}$, the swept band is $\Delta\Phi_{\rm sweep}\approx96^\circ$; with
half-widths $\omega=20^\circ$--$45^\circ$ for large events, Eq.~\eqref{eq:ratio} gives
$\nu_{\rm conn}/\nu_{\rm impact}\approx2$--$3.5$. A fast CME that stays super-critical to
$r_a\sim0.7\unit{AU}$ widens the band to $\sim140^\circ$ and the ratio to $\sim4$; conversely,
the observed fall-off of gradual-event intensity with connection angle ($e$-folding
$\sim40^\circ$) down-weights flank connections, returning a fluence-weighted ratio near
$2$--$3$.

Read this way, the nominal forcing rate rises from an impact value $\nu_{\rm impact}\sim10^{2}$
to a connection value $\nu_{\rm conn}\sim2.5$--$3\times10^{2}\unit{yr^{-1}}$, lifting
Eq.~\eqref{eq:globalN} to
\begin{equation}
  \dot N_{\mathrm{fix}}\;\sim\;(4\e{10}\;\text{--}\;4\e{11})\unit{mol\,N\,yr^{-1}}
  \label{eq:globalNconn}
\end{equation}
This moves the operating point up the $\nu$ axis of the regime map (Fig.~\ref{fig:regime}) by
$\sim0.5$~dex, well inside the prebiotic-pool region but still nearly an order of magnitude below
FYS closure, without requiring a harder spectrum. The connection rate declines with age in the
same way as the total CME rate, so the self-extinguishing thermostat (\S\ref{sec:thermostat})
keeps its shape and is only rescaled in amplitude. The chemically important region of the
$(E_b,\nu)$ plane is therefore reached at a CME occurrence rate several times lower than an
impact-based estimate would require.

\section{The Alfv\'en surface: from the modern to the young Sun}
\label{sec:alfven}

\paragraph{Why it matters.}
The Alfv\'en surface is the locus where the radial wind speed first equals the local Alfv\'en
speed, $v(r)=v_A(r)$. Inside it the magnetic field dominates the plasma: the corona co-rotates
with the star, disturbances can propagate back toward the Sun, and the wind remains in causal
contact with its source; outside it the flow is inertially dominated, escapes irreversibly, and
combs the field into the Parker spiral. The size of this surface sets the lever arm for
angular-momentum loss and stellar spin-down \citep{reville2015,finley2018}, anchors the base of
the heliospheric field, and defines the magnetic ``cage'' within
which eruptions and flare-accelerated particles are confined near the Sun. Parker Solar Probe
now crosses the surface routinely \citep{kasper2021,finley2025}, so the modern value is a
measured quantity. Because the young Sun's field was far stronger, its Alfv\'en surface
stood much farther out, with direct consequences for CME escape and SEP transport.

\paragraph{Calculation.}
We build each epoch's radial profiles from conservation laws, following the framework used in
PSP analyses \citep{finley2025}. The Alfv\'en speed is
\begin{equation}
  v_A(r) = \frac{B(r)}{\sqrt{\mu_0\,\rho(r)}} .
  \label{eq:vA}
\end{equation}
Conservation of magnetic flux and of mass flux along a steady, radially expanding flux tube give
\begin{equation}
  B(r)\,r^2=\mathrm{const}\;\Rightarrow\; B(r)=B_0\!\left(\frac{R_\odot}{r}\right)^{2}\!,
  \qquad
  \rho(r)\,v(r)\,r^2=\frac{\dot M}{4\pi}\;\Rightarrow\;
  \rho(r)=\frac{\dot M}{4\pi r^2 v(r)} ,
  \label{eq:flux}
\end{equation}
with $B_0$ the surface radial field and $\dot M$ the mass-loss rate. The wind accelerates from
the base to a terminal speed $v_\infty$,
\begin{equation}
  v(r)=v_0+(v_\infty-v_0)\left[1-\exp\!\big(-(r-R_\odot)/h_v\big)\right],
  \label{eq:vwind}
\end{equation}
and the Alfv\'en radius $r_A$ is the root of $v(r_A)=v_A(r_A)$. Inserting Eq.~\eqref{eq:flux}
into \eqref{eq:vA} gives $v_A(r)=(B_0R_\odot^2/r)\sqrt{4\pi v(r)/(\mu_0\dot M)}$, so up to the
weak $\sqrt{v}$ factor the surface obeys
\begin{equation}
  r_A\;\propto\;\frac{B_0}{\sqrt{\dot M\,v_\infty}}\;\sim\;\Upsilon^{\,m},
  \qquad
  \Upsilon=\frac{\phi_{\rm open}^{2}}{R_\odot^{2}\,\dot M\,v_{\rm esc}} ,
  \label{eq:rAscale}
\end{equation}
the familiar dependence of the Alfv\'en radius on the wind magnetisation $\Upsilon$
\citep{reville2015,finley2018}: a stronger field pushes the surface out, while a higher mass
loss or faster wind pulls it in.

\begin{figure}[t]
\centering
\includegraphics[width=0.8\textwidth]{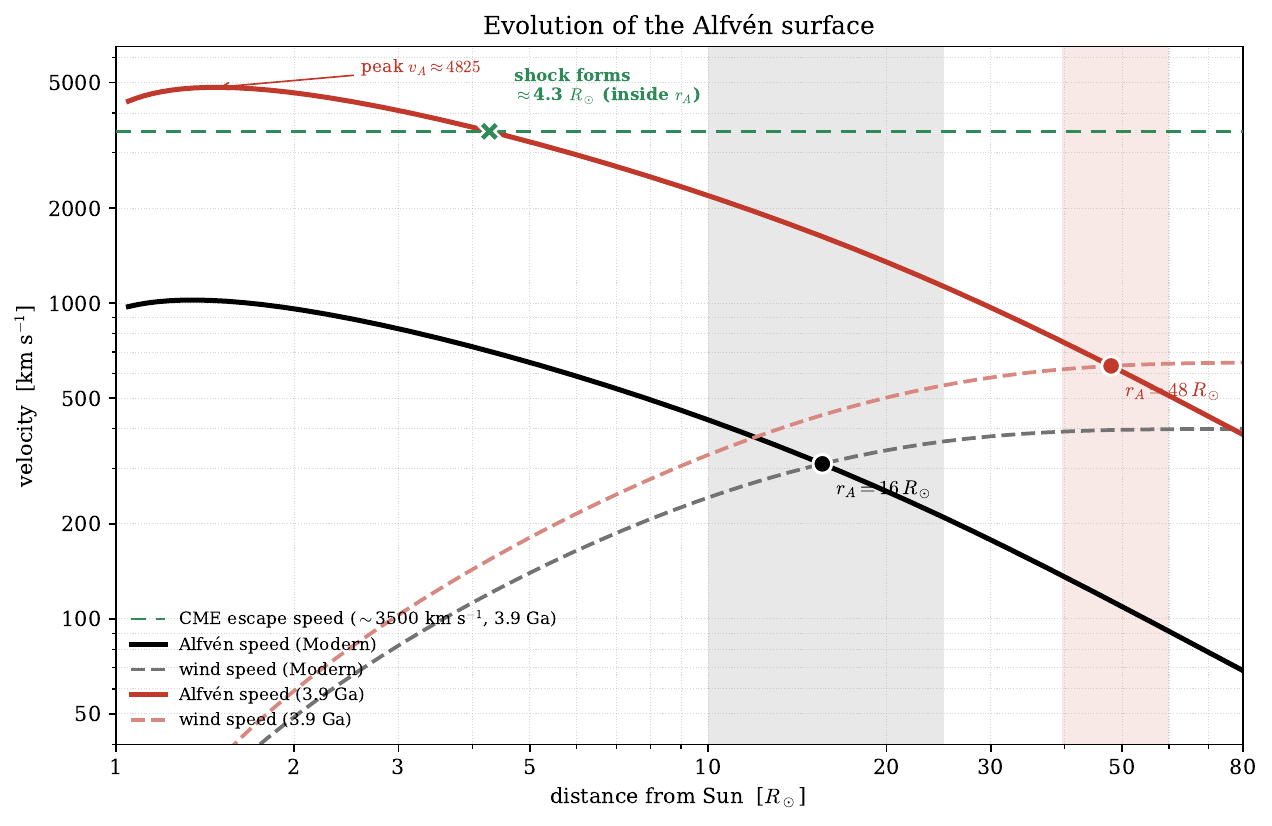}
\caption{Evolution of the Alfv\'en surface (log--log). Solid curves are the Alfv\'en speed
$v_A=B/\sqrt{\mu_0\rho}$ (Eq.~\ref{eq:vA}), dashed curves the wind speed $v(r)$
(Eq.~\ref{eq:vwind}), for the modern Sun (black) and 3.9~Ga (red); filled dots mark the wind
Alfv\'en points $v(r_A)=v_A(r_A)$ at $r_A\approx16$ and $48\,R_\odot$. The bands give the
corresponding ranges of surface height: grey from PSP observations
($\sim$$10$--$25\,R_\odot$; cycle-averaged $11$--$19\,R_\odot$;
\citealp{finley2025,liu2023,kasper2021}) and red from the model ($\sim$$40$--$60\,R_\odot$). The
surface scales as $r_A\propto B_0/\sqrt{\dot M\,v_\infty}$ (Eq.~\ref{eq:rAscale}). The green dashed line is the $\sim$3500$\unit{km\,s^{-1}}$ an ejection
must reach to escape at 3.9~Ga (\S\ref{sec:decel}). Because the strong field makes $v_A$ peak at
$\sim$4800$\unit{km\,s^{-1}}$ near $1.5\,R_\odot$, this speed lies below the peak: the
ejection is trans-Alfv\'enic at launch and its shock forms only where $v_A$ falls to meet it
(green cross, $\approx$$4\,R_\odot$), deep inside the wind Alfv\'en surface. The CME
shock radius (set by the fast ejecta) and the wind Alfv\'en surface (set by the slow wind) are
thus distinct crossings, separated by an order of magnitude in radius.}
\label{fig:alfven}
\end{figure}

\paragraph{Modern calibration.}
For the present Sun the surface is measured. PSP places the cycle-averaged Alfv\'en
radius at $11$--$19\,R_\odot$ (smallest at solar minimum, largest in the rising phase), with the
instantaneous, rugged surface fluctuating by $10$--$40\%$ in space and time and spanning
$\sim$$10$--$30\,R_\odot$ \citep{finley2025,liu2023}; independent PSP extrapolations give
$8$--$19\,R_\odot$ \citep{cranmer2023}, consistent with the broader $10$--$40\,R_\odot$ in-situ
and $10$--$20\,R_\odot$ remote-sensing bounds \citep{verscharen2021,deforest2014}. The first
direct crossing was at $18.8\,R_\odot$ \citep{kasper2021}. We anchor the modern band in
Fig.~\ref{fig:alfven} to the observed $\sim$$10$--$25\,R_\odot$ range.

\paragraph{Result.}
Figure~\ref{fig:alfven} compares the present Sun with the 3.9~Ga configuration, using
$(v_\infty,B_0,\dot M/\dot M_\odot)=(400\unit{km\,s^{-1}},1.4\unit{G},1)$ today and
$(650\unit{km\,s^{-1}},55\unit{G},80)$ at 3.9~Ga, where the field is $\sim$30$\times$ stronger
(\S\ref{sec:bc}) and the mass loss follows the $\dot M\propto t^{-2.3}$ trend
\citep{wood2005}. The Alfv\'en surface moves from $r_A\approx16\,R_\odot$ (observed band
$10$--$25$) to $r_A\approx48\,R_\odot$ (model band $40$--$60$) at 3.9~Ga, an outward expansion
by a factor of $\sim$3. The young-Sun band reflects the uncertain mass-loss history; the modern
band is observational.

The moderate wind speed of the active-corona topology (\S\ref{sec:environment}) also helps keep
the surface at $40$--$60\,\Rsun$. Since $r_A\propto B_0/\sqrt{\dot M\,v_\infty}$
(Eq.~\ref{eq:rAscale}), the uniformly fast ($\gtrsim$1500$\unit{km\,s^{-1}}$) wind sometimes
assumed for active stars would draw $r_A$ inward by $\sim\!\sqrt{v_{\rm fast}/v_\infty}$, a factor
of $\sim$1.6, and partly offset the field-driven expansion. The Alfv\'enic slow wind is therefore
part of the reason the young-Sun Alfv\'en surface lies so far out.

\paragraph{Consequences: CME caging, failed eruptions, and impulsive SEPs.}
An Alfv\'en surface at $\sim$50$\,R_\odot$ enlarges threefold in radius the region where the
field dominates the plasma ($v<v_A$, low $\beta$). Three consequences follow.

\textit{CME caging.} An ejection must overcome the tension and ram pressure of the overlying
field to escape, and that confining field now reaches three times farther out and is
$\sim$30$\times$ stronger. A larger fraction of eruptions, particularly the slower and less
massive ones, is decelerated and contained before reaching the super-Alfv\'enic wind, so the
escaping CME fraction falls while the trapped energy is redeposited in the inner corona.
Only the fastest, most energetic ejections break through cleanly; since these also drive the
strong shocks that make gradual SEPs, the cage works like a high-pass filter that favours the
SEP-productive events.

\textit{Failed eruptions.} In the same strong-field regime the torus and kink instabilities that
release flux ropes reach their thresholds (e.g.\ the critical field-decay index) only higher in
the corona, so a larger share of eruptions stall as failed, confined events that rise and are then
held down by the overlying field, as is often inferred for magnetically active young stars. This throttles the low-energy tail of the CME distribution and diverts its energy into
repeated confined reconnection and heating rather than into escaping ejecta.

\textit{Impulsive SEP propagation.} The enlarged Alfv\'en surface also enlarges the magnetic
bottle for flare-accelerated particles. Impulsive SEPs injected low in the corona must climb out
through a mirror that is stronger and three times more extended, so a smaller fraction escapes to
1~AU; \S\ref{sec:transport} quantifies this escape fraction $f_{\rm esc}$. The gradual channel
behaves differently, and the reason requires distinguishing two Alfv\'en conditions. The
Alfv\'en surface, $r_A\!\approx\!48\,R_\odot$, is set by the slow background wind
($\sim$650$\unit{km\,s^{-1}}$). The radius at which a CME drives a shock is set by the much faster
ejecta and lies far deeper. The strong young-Sun field makes the Alfv\'en speed peak at
$v_A\!\sim\!4800\unit{km\,s^{-1}}$ near $1.5\,R_\odot$ and fall outward. An ejection is therefore
trans-Alfv\'enic at launch, which is why it must reach $\sim$3500--4000$\unit{km\,s^{-1}}$ to
break out (\S\ref{sec:decel}). It becomes super-Alfv\'enic and drives a strong shock only where
$v_A$ falls to meet it, near $r\!\approx\!5\,R_\odot$, still well inside $r_A$. From there the
shock is a continuous, outward-moving source that stays super-Alfv\'enic through $r_A$ and far
beyond. Most of its accelerating path, and hence most of its fluence, lies at or beyond $r_A$,
where escape is free, and even its inner injections occur high in the cage, far above the flare
site. The same strong field thus sets the CME breakout threshold (close to the peak coronal
$v_A$) and the confinement penalty for flare particles, while the gradual channel is comparatively
unimpeded. Section~\ref{sec:transport} quantifies this asymmetry.

\section{Transport and decay of young-Sun SEPs: impulsive versus CME-driven}
\label{sec:transport}

The two SEP classes introduced in \S\ref{sec:intro}, impulsive (flare and jet) and gradual
(CME shock), differ in their sources and in how their fluence is delivered. The decay phase of a
gradual event is governed mainly by interplanetary transport: field-aligned diffusion,
convection, and adiabatic deceleration in the expanding wind \citep{parker1965,desai2016}. At late
times this produces the familiar, spatially near-uniform ``reservoir'' of slowly fading
intensity. These properties are calibrated on the modern heliosphere. At 3.9~Ga the stronger
field, tighter spiral, denser wind, and enlarged Alfv\'en surface change every transport
parameter at once, and they push the two channels in opposite directions. We quantify the
consequences below.

The atmospheric forcing depends on the time-integrated SEP flux (the fluence) delivered near
1~AU over the whole event. The fluence is set jointly by how fast an event decays and by how long
its source keeps injecting. We compare the two channels
with a one-dimensional, field-aligned transport estimate, writing the impulsive channel as a
single near-Sun injection and the gradual (CME ``pile-up'') channel as an extended, moving
shock source.

Along the flux tube the parallel diffusion coefficient is
\begin{equation}
  \kappa_\parallel = \tfrac{1}{3}\, v\, \lambda_\parallel ,
  \label{eq:kappa}
\end{equation}
with $v$ the particle speed and $\lambda_\parallel$ the parallel mean free path. The competing
loss timescales at heliocentric distance $r$ are
\begin{equation}
  \tau_\parallel = \frac{L^2}{2\kappa_\parallel},\qquad
  \tau_c = \frac{r}{V_{\rm sw}},\qquad
  \tau_{\rm ad} \approx \frac{3}{2}\,\frac{r}{V_{\rm sw}},\qquad
  \tau_\perp = \frac{(\Delta r)^2}{\kappa_\perp},\quad \frac{\kappa_\perp}{\kappa_\parallel}\sim0.01\text{--}0.05 ,
  \label{eq:taus}
\end{equation}
for parallel diffusion along a field line of length $L$, convection past the observer,
adiabatic deceleration out of a fixed energy band, and cross-field (``cage'') leakage. The
decay is governed by the fastest loss; we lump convection, adiabatic cooling, and leakage into
an effective $\tau_{\rm loss}$. For an impulsive injection a field-line distance $L$ away, the
first-passage (inverse-Gaussian) intensity with catastrophic loss is
\begin{equation}
  G(t;L,\kappa_\parallel,\tau_{\rm loss}) =
  \frac{L}{\sqrt{4\pi\,\kappa_\parallel\,t^{3}}}\,
  \exp\!\left(-\frac{L^{2}}{4\kappa_\parallel t}\right)\,
  \exp\!\left(-\frac{t}{\tau_{\rm loss}}\right),
  \qquad t_{\rm peak}=\frac{L^{2}}{6\kappa_\parallel}.
  \label{eq:green}
\end{equation}
The impulsive channel is a single injection delivered to 1~AU through the near-Sun magnetic
bottle with escape fraction $f_{\rm esc}$, while the gradual channel is an extended,
moving source: the shock injects continuously from where it first becomes
super-magnetosonic (low in the corona, $r\!\sim\!5\,R_\odot$, well inside the wind Alfv\'en
radius $r_A$; \S\ref{sec:alfven}) until it weakens, and particles
launched at time $t'$ traverse only the remaining path $L(t')$ to 1~AU,
\begin{align}
  I_{\rm flare}(t) &= A\, f_{\rm esc}\, G(t;L,\kappa_\parallel,\tau_{\rm loss}), \label{eq:Iflare}\\
  I_{\rm CME}(t)   &= A\!\int_{0}^{t_{\rm inj}}\! q(t')\,
     G\!\big(t-t';\,L(t'),\kappa_\parallel,\tau_{\rm loss}\big)\,dt',\quad
     L(t') = L\big[1-(1-f_{R})\,t'/t_{\rm inj}\big], \label{eq:Icme}
\end{align}
with $q(t')$ the injection rate ($\propto$ shock strength), $t_{\rm inj}$ the strong-shock
duration, and $f_{R}$ the residual path fraction when the shock nears 1~AU. The delivered
fluence is $F=\int I\,dt$.

Four young-Sun levers enter. (i)~The tight Parker spiral lengthens the field-line path to
$L\approx2.2\unit{AU}$ (vs.\ $\approx1.2\unit{AU}$ today; \S\ref{sec:spiral}), raising
$\tau_\parallel\propto L^{2}$. (ii)~The denser, more turbulent wind shortens
$\lambda_\parallel$ ($\approx0.07$ vs.\ $\approx0.2\unit{AU}$), lowering $\kappa_\parallel$.
(iii)~The faster wind ($V_{\rm sw}\approx650$ vs.\ $400\unit{km\,s^{-1}}$) shortens $\tau_c$,
but the stronger field suppresses $\kappa_\perp$ and lengthens $\tau_\perp$; the net is a
modestly longer $\tau_{\rm loss}$. (iv)~The Alfv\'en surface, pushed to
$R_{\rm A}\approx40\text{--}60\,R_\odot$ by the $\sim$30$\times$ field,
\begin{equation}
  V_{\rm sw}(R_{\rm A}) = V_{\rm A}(R_{\rm A}),\qquad V_{\rm A}=\frac{B}{\sqrt{\mu_0\rho}},
  \label{eq:RA}
\end{equation}
plays an asymmetric role: it enlarges the magnetic bottle that confines flare particles,
suppressing their escape fraction $f_{\rm esc}$, while the CME shock turns on only near/beyond
$R_{\rm A}$, so the gradual particles are born outside the bottle and escape freely
($f_{\rm esc}\!\approx\!1$). The same enlarged $R_{\rm A}$ thus chokes the impulsive channel and
frees the gradual one.

\begin{table}[h]
\centering
\caption{Transport parameters and outcomes for the two channels (Eqs.~\ref{eq:kappa}--\ref{eq:Icme})
for a $\sim$10~MeV proton.}
\label{tab:transport}
\begin{tabular}{lcc}
\toprule
quantity & today & 3.9~Ga \\
\midrule
field-line path $L$ [AU]                 & 1.2 & 2.2 \\
mean free path $\lambda_\parallel$ [AU]  & 0.20 & 0.07 \\
$\kappa_\parallel$ [AU$^2$\,d$^{-1}$]    & 1.8 & 0.6 \\
effective $\tau_{\rm loss}$ [d]          & 2.0 & 3.5 \\
flare escape fraction $f_{\rm esc}$      & 1.0 & 0.22 \\
CME injection $t_{\rm inj}$ [d]          & 0.8 & 2.0 \\
\midrule
decay e-folding $\tau_{\rm decay}$ (flare / CME) [d] & 0.6 / 0.5 & 2.1 / 1.3 \\
fluence ratio CME:flare                  & ... & $\approx$24 : 1 \\
\bottomrule
\end{tabular}
\end{table}

\begin{figure*}[]
\centering
\includegraphics[width=\textwidth]{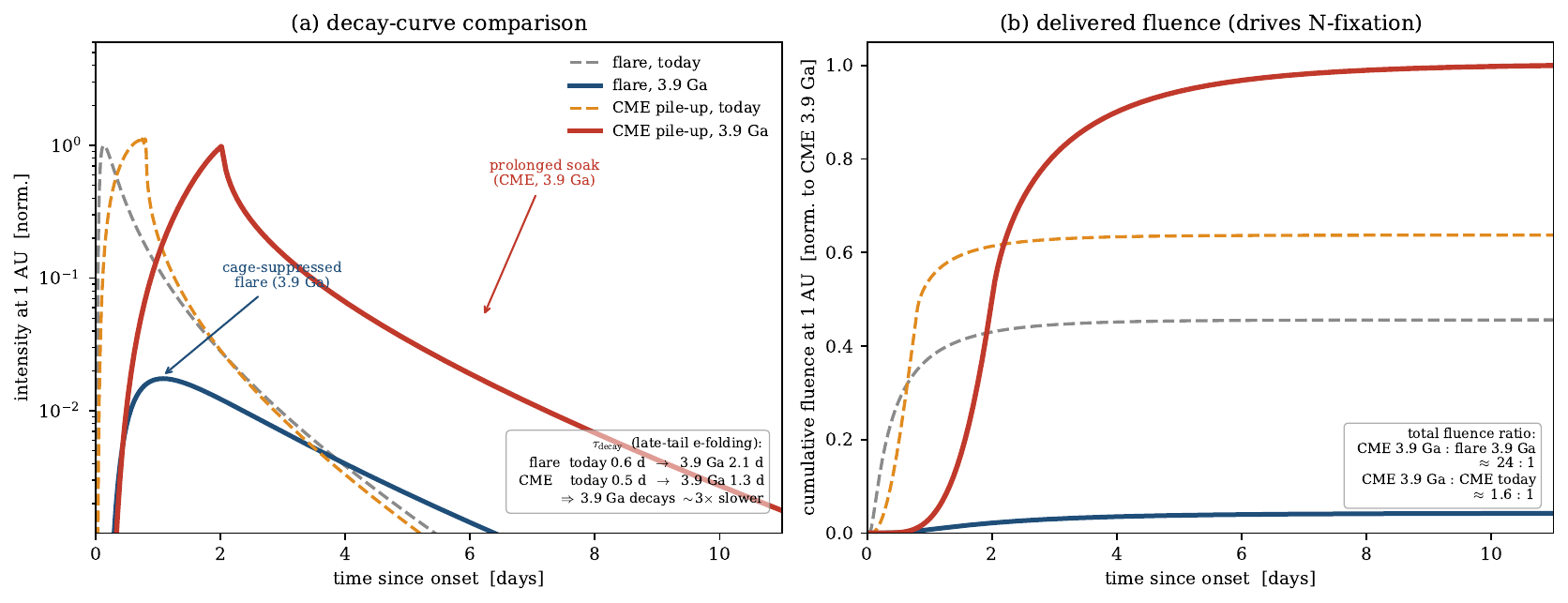}
\caption{SEP intensity-time profiles at 1~AU \textbf{(a)} and the cumulative fluence they
deliver \textbf{(b)}, for the impulsive (flare) and moving CME-shock (pile-up) channels under
modern and 3.9~Ga transport (Table~\ref{tab:transport}, Eqs.~\ref{eq:kappa}--\ref{eq:Icme}).
The late-tail decay e-folding lengthens by $\sim$3$\times$ from today to 3.9~Ga; the
CME-pile-up channel at 3.9~Ga delivers $\sim$24$\times$ the integrated fluence of the
cage-suppressed impulsive channel and $\sim$1.6$\times$ the modern CME event.}
\label{fig:sepdecay}
\end{figure*}

Figure~\ref{fig:sepdecay} shows the result. The late-time decay e-folding lengthens from
$\tau_{\rm decay}\approx0.5\text{--}0.6\unit{d}$ today to $\approx1.3\text{--}2.1\unit{d}$ at
3.9~Ga. This factor of $\sim$3 slower decay is set mainly by the longer path and shorter
$\lambda_\parallel$ and is only partly offset by the faster wind. For the chemistry the fluence
matters more. At 3.9~Ga the CME pile-up channel delivers $\sim$24 times the integrated fluence of
the impulsive channel, which the enlarged cage suppresses, and $\sim$1.6 times that of a modern
CME event. The prolonged particle exposure invoked for the greenhouse and thermostat arguments
(\S\ref{sec:thermostat}) is therefore the convolution of a multi-day moving-shock injection with
this slower, tightly trapped decay, while the strong field suppresses the impulsive background.
The main uncertainties are $\lambda_\parallel$, because the young-Sun turbulence spectrum is
unmeasured, and $f_{\rm esc}$, which needs a proper loss-cone and mirror calculation. The
$\sim$3$\times$ slowing depends on the former and the $24\!:\!1$ ratio partly on the latter, so
both are scaling estimates. A focused-transport integration (e.g.\ \textsc{eprem},
\textsc{ipath}) with young-Sun $B$, $\lambda_\parallel$, and a moving shock source, coupled to the
EUHFORIA run of \S\ref{sec:decel}, would turn these scalings into firm curves.

\paragraph{From channel asymmetry to the fixation rate.}
This asymmetry justifies using a single channel downstream. The fixation rate of
\S\ref{sec:fys} (Eq.~\ref{eq:globalN}) is built from the CME-shock channel alone. With the
$\sim$$24\!:\!1$ dominance, the impulsive channel adds only $\sim$4\% to the integrated fluence,
so neglecting it is a controlled approximation. The enlarged Alfv\'en cage (\S\ref{sec:alfven})
is the reason the gradual channel sets the chemistry: it suppresses the flare contribution and
lengthens the gradual exposure that builds the long-lived $\mathrm{N_2O}$ reservoir. The cage and
transport sections therefore enter the final estimates through a single number, the
gradual-channel fluence, with the channel ratio entering only as a few-percent correction.

\subsection{The particle--UV decoupling}
\label{sec:uvdecouple}

A second decoupling, distinct from the connection-versus-impact geometry of
\S\ref{sec:decouple}, separates the particle forcing from the ultraviolet (UV) that accompanies
it. SEPs are charged particles and radiate negligible UV directly. The UV associated with an event
comes from the parent flare's prompt electromagnetic emission and from secondary airglow, excited
when the precipitating particles collide with $\mathrm{N_2}$ and $\mathrm{O_2}$. These sources
weight the two channels oppositely. Impulsive events share a reconnection origin with a strong,
prompt UV/EUV burst, so their particles and UV are tightly correlated. The gradual channel is
accelerated at a CME-driven shock, which is an efficient particle accelerator but a poor UV
source: it produces an EUV wave and coronal dimming, with no flash. Its accompanying UV comes from
the associated flare, if any, and tracks the gradual fluence only loosely. The gradual channel
dominates the particle budget (by $\sim$24:1; \S\ref{sec:transport}), whereas flares dominate the
prompt UV, so the nitrogen-fixing fluence and the destructive UV burst are partly independent.
Particles can also change the surface UV indirectly, by altering the atmospheric UV shield. In the
present ozone-bearing atmosphere, energetic particles deplete ozone, and the highest surface UV-B
yet recorded, a UV index of 43.3 in the Bolivian Andes, followed a large solar flare by about two
weeks, although the connection remains circumstantial \citep{cabrol2014}. In the anoxic Archean
atmosphere the corresponding shield is the organic haze discussed below.

Scaling estimates make the separation concrete. Following the flare/CME energy partition
\citep{emslie2012}, the flare's radiated energy is comparable to the CME kinetic energy,
$\Delta E_k\sim6\e{25}\unit{J}$, and its UV/EUV portion is of order $10$--$40\%$, i.e.\
$\sim$$10^{24}$--$10^{25}\unit{J}$ per event. This is comparable to or larger than the SEP energy
itself ($\Esep\sim6\e{24}\unit{J}$). The secondary airglow is far smaller. With
$E_{\rm dep}\sim10^{14}$--$10^{16}\unit{J}$ deposited below the survival level
(\S\ref{sec:sepprod}) and about a percent re-radiated in the UV, it contributes
$\sim$$10^{12}$--$10^{14}\unit{J}$ per event, some ten orders of magnitude below the flare term.
The flare therefore dominates the UV energy, but the airglow is co-located with the precipitation
and is present on the night side and beneath haze, where direct solar UV is absent.

The decoupling matters for the downstream chemistry in two ways. For the $\mathrm{N_2O}$ budget
(\S\ref{sec:fys}), UV near $\sim$$180$--$220\unit{nm}$ is the photolytic destruction term that the
haze shielding offsets. Transient flare bursts raise the instantaneous loss but are brief, so the
steady-state lifetime, and hence the net forcing of \S\ref{sec:netforcing}, is set by the mean UV
field together with the haze. For prebiotic chemistry, UV in the $\sim$$200$--$280\unit{nm}$ band
is an energy source as well as a hazard \citep{ranjan2016,rimmer2018,patel2015}. The dominant
fixation channel (gradual) therefore delivers nitrogen feedstock with comparatively little
coincident hard UV, while impulsive events bring strong prompt UV that both drives and degrades
the chemistry. The haze reinforces this decoupling. A fractal aggregate haze absorbs strongly in
the ultraviolet and so shields the photolabile species beneath it; \citet{wolftoon2010} showed
that the same shield that warms the surface protects gases such as $\mathrm{NH_3}$ from
photolysis. In the hazy (high-$\mathrm{CH_4}$, biologically active) regime, the UV that would
otherwise destroy the particle-produced $\mathrm{N_2O}$ is therefore attenuated. This partly
decouples the $\mathrm{N_2O}$ lifetime and the survival of prebiotic feedstock from the
instantaneous UV field. The fractal limit that minimises the antigreenhouse penalty
(\S\ref{sec:netforcing}) also preserves the particle-channel $\mathrm{N_2O}$ and so strengthens
the net-positive forcing.

Two caveats bound this picture. First, the fastest CMEs, which make the largest gradual events,
usually do carry an associated flare, so the tendency of gradual events to be UV-quiet is only
statistical. Second, a quantitative treatment requires folding the flare UV
spectrum against the $\mathrm{N_2O}$ photolysis cross-section and the prebiotic action spectrum
\citep{segura2010,ranjan2016,rimmer2018}; we therefore present the particle--UV decoupling as a
qualitative organizing principle and a target for the coupled photochemical--radiative modelling
for future work.

\section{Influence on the Faint Young Sun paradox}
\label{sec:fys}

The FYS requirement is twofold. At 3.9~Ga, an enhanced greenhouse must supply
$\Delta F\sim 20\text{--}40\unit{W\,m^{-2}}$, the part of the $\approx60\unit{W\,m^{-2}}$
deficit that cloud and surface-albedo changes cannot cover (\S\ref{sec:intro}). That forcing
must also fade as $L_\odot$ rises, so that the planet does not overheat by $\sim 2\unit{Ga}$.

Beyond the broad greenhouse/albedo division noted in \S\ref{sec:intro}, the specific greenhouse
candidates include elevated $\mathrm{CO_2}$ \citep{kasting1993}, biological or abiotic
$\mathrm{CH_4}$ with organic-haze feedbacks \citep{pavlov2000,haqqmisra2008}, and
collision-induced absorption by $\mathrm{H_2}$--$\mathrm{N_2}$ \citep{wordsworth2013}. Fractal
organic haze can also shield reduced greenhouse gases from photolysis while cooling only weakly
\citep{wolftoon2010,wolftoon2013}. \citet{airapetian2016} introduced a space-weather pathway: SEPs
from frequent young-Sun CMEs fix atmospheric nitrogen into $\mathrm{N_2O}$, a greenhouse gas
$\sim 300\times$ more potent than $\mathrm{CO_2}$ per molecule, which both warms the planet and
seeds prebiotic chemistry. Later magnetospheric modelling \citep{sur2025} and the EK~Dra CME
detection \citep{namekata2025} strengthen the empirical case for sustained, intense forcing. We
extend this picture by sourcing the SEPs from heliospheric deceleration, so that the forcing
inherits the measured $\dot M\propto t^{-2.3}$ wind history \citep{wood2005} and declines with
time. We also show that the sign of the net radiative effect depends on the SEP spectral hardness
(the break energy $E_b$), and we quantify the radiative amplification the chemistry must supply.

\subsection{Direct SEP heating is negligible}

The time-averaged SEP power into the atmosphere is negligible as direct heat:
\begin{equation}
  P_{\mathrm{SEP}} = \frac{E_{\mathrm{atm}}\,\nu}{A_\oplus}
  = \frac{1.6\e{16}\times 100}{5.1\e{14}\times 3.15\e{7}}
  \approx 1\e{-4}\unit{W\,m^{-2}},
\end{equation}
($\nu=100\unit{yr^{-1}}$). Any climatic effect must therefore come through radiatively active
chemistry. As an order-of-magnitude estimate, the chemistry must provide an amplification
\begin{equation}
  \mathcal{L} = \frac{\Delta F}{P_{\mathrm{SEP}}}
  \sim \frac{30}{10^{-4}} \sim 3\e{5}.
  \label{eq:leverage}
\end{equation}
Each watt of particle energy must control $\sim 10^5\unit{W}$ of greenhouse forcing. This is
possible only because a long-lived trace gas, made at a cost of $\sim 100\unit{eV}$ per molecule,
then absorbs infrared radiation for decades. This amplification works only if the molecule forms deep
enough to survive, so it is gated by $\fdep(p_s;E_b)$ from Eq.~\eqref{eq:fpen}. Hard spectra
deposit their energy below the survival level, where the products can persist. Soft spectra make
NO high in the atmosphere, where it is photolysed and destroyed.

\subsection{The greenhouse budget and a conservative bound}
\label{sec:budget}

To hold $\mathrm{N_2O}$ at $\sim 1\unit{ppm}$ ($N\sim 1.7\e{14}\unit{mol}$) with an anoxic-atmosphere
lifetime $\tau\sim 50\unit{yr}$ requires
\begin{equation}
  P_{\mathrm{N_2O}} = N/\tau \sim 3\e{12}\unit{mol\,yr^{-1}},
\end{equation}
whereas Eq.~\eqref{eq:globalN} supplies $1.4\e{10}\text{--}1.4\e{11}\unit{mol\,yr^{-1}}$ at the
impact rate, or $\sim3\times$ more once $\nu$ is read as the connection rate
(Eq.~\eqref{eq:globalNconn}, \S\ref{sec:decouple}). Two separate conditions must be met. The
first is deposition below the survival level $p_s$, which requires a hard spectrum. Such spectra
are readily available. Efficient near-Bohm acceleration at the strong, turbulence-driving
young-Sun shock places the break as high as $E_b\sim100\unit{MeV}$, compared with the
conservative $E_b\sim30\unit{MeV}$ of Table~\ref{tab:fpen}. This raises the fraction of SEP energy
deposited below 3~hPa ($\approx$40~km today) from $\sim4\%$ to $\sim30\%$, and the
$>100\unit{MeV}$ tail is populated in any case. The strong shock implied by the deceleration,
threading the strong young-Sun field, if anything favours high-energy acceleration. The second
condition is abundance, and this is where the shortfall lies. Even with deep deposition and the
connection rate, the steady-state $\mathrm{N_2O}$ stays more than two orders of magnitude below
$1\unit{ppm}$ at the canonical $\tau\sim50\unit{yr}$. Closing the gap requires a longer
$\mathrm{N_2O}$ lifetime, which fractal organic haze can provide by shielding $\mathrm{N_2O}$ from
photolysis and raising $\tau$ by one to two orders of magnitude (\S\ref{sec:netforcing}). The
connection-rate frequency and, plausibly, larger event masses would also be needed. The closure
is therefore marginal and depends on where the system sits in parameter space. That position is
set by the photochemical lifetime and the event statistics; the shock itself has no difficulty
producing energetic particles.

\subsection{The net forcing, including the haze's own cooling}
\label{sec:netforcing}

\begin{figure}[h]
\centering
\includegraphics[width=0.8\textwidth]{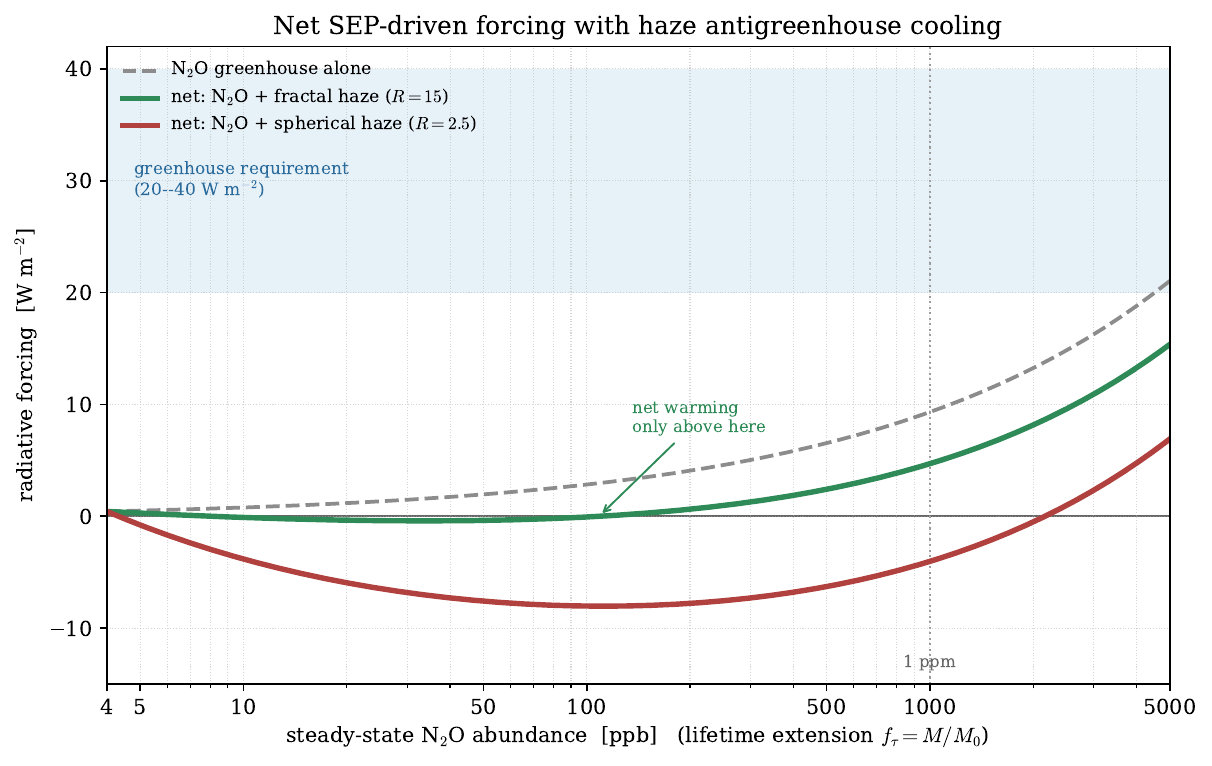}
\caption{Net SEP-driven radiative forcing once the haze's own antigreenhouse cooling is
included. The $\mathrm{N_2O}$ greenhouse alone (grey dashed) grows with abundance, but realising
that abundance requires extending the $\mathrm{N_2O}$ lifetime by $f_\tau=M/M_0$ through
ultraviolet shielding, and the shielding haze cools the surface. The net forcing is positive, a
few $\unit{W\,m^{-2}}$ and a partial contribution to the 20--40$\unit{W\,m^{-2}}$ greenhouse
requirement of \S\ref{sec:fys} (blue band), only for fractal haze (green; high
ultraviolet/visible opacity ratio $R$). Spherical haze (red) cancels the warming, so closure
depends on haze morphology as well as on particle energy and abundance. Curves are an
order-of-magnitude band model (\S\ref{sec:netforcing}).}
\label{fig:netforcing}
\end{figure}

The lifetime extension invoked above has a cost. The fractal organic haze that shields
$\mathrm{N_2O}$ from photolysis is itself radiatively active, and its antigreenhouse cooling
must be subtracted before any warming can be claimed. We make this explicit with an
order-of-magnitude budget (Fig.~\ref{fig:netforcing}), a scaling estimate that does not replace a
radiative transfer calculation. Writing the steady-state abundance as $M=M_0\,f_\tau$ with
$M_0\sim4\unit{ppb}$ the no-haze (connection-rate) value and $f_\tau$ the lifetime-extension
factor, the $\mathrm{N_2O}$ forcing follows a band-saturated form
$\Delta F_{\mathrm{N_2O}}\simeq c_N(\sqrt{M}-\sqrt{M_{\rm bg}})$. The same shielding that yields
$f_\tau$ requires a haze ultraviolet optical depth $\tau_{\rm UV}=\ln f_\tau$ (photolysis rate
$\propto e^{-\tau_{\rm UV}}$), to which corresponds a visible optical depth
$\tau_{\rm vis}=\tau_{\rm UV}/R$ and a cooling
$\Delta F_{\rm haze}\simeq-k_h\,(1-e^{-\tau_{\rm vis}})$, where $R$ is the ultraviolet-to-visible
opacity ratio. Haze morphology enters through $R$ alone: fractal aggregates extinguish the
ultraviolet efficiently while remaining nearly transparent in the visible ($R\sim15$), whereas
compact spheres do not ($R\sim2$--3) \citep{wolftoon2010}.

The net forcing $\Delta F_{\rm net}=\Delta F_{\mathrm{N_2O}}+\Delta F_{\rm haze}$ then depends
strongly on morphology. For fractal haze the cooling stays small ($\tau_{\rm vis}\lesssim0.5$ even
at $f_\tau\sim10^3$), and the net turns positive above $M\sim0.1\unit{ppm}$. It reaches a few
$\unit{W\,m^{-2}}$ at ppm-level $\mathrm{N_2O}$, a meaningful fraction of the
$20$--$40\unit{W\,m^{-2}}$ requirement but not enough to close it. For spherical haze the visible
optical depth grows with the shielding, so the cooling cancels the greenhouse and the net stays
negative across the plausible range. The SEP--$\mathrm{N_2O}$ term therefore warms the planet only
in the fractal-haze regime. Its climatic relevance depends on an aerosol property that
photochemical-aerosol models, and eventually the Archean rock record, can constrain.

$\mathrm{N_2O}$ does not carry the SEP-driven greenhouse budget alone. In the coupled
photochemical--climate chains of \citet{biasiotti2026}, the species that moves the surface
temperature is CO, which reaches $\sim$$70$--$150\unit{ppm}$ in the processed atmosphere and
accounts for nearly all of their $\sim$$0.3\unit{K}$ warming. $\mathrm{H_2}$ contributes where it
survives, through $\mathrm{CO_2}$--$\mathrm{H_2}$ collision-induced absorption
\citep{pierrehumbert2011}, and the nitrogen species ($\mathrm{N_2O}$, HCN) remain minor. Our
deceleration-powered pulse processes the same $\mathrm{N_2}$-dominated column, so it modulates
this $\mathrm{CO}/\mathrm{H_2}$ channel together with $\mathrm{N_2O}$. The $\mathrm{CO}/\mathrm{H_2}$
contribution is not limited by the UV photolysis that limits $\mathrm{N_2O}$, and it does not
depend on haze morphology. It therefore broadens the transient forcing and reduces the mechanism's
reliance on $\mathrm{N_2O}$, the most contested species. It also persists in the spherical-haze
case, where the $\mathrm{N_2O}$ term turns net-negative.

\begin{figure}[h]
\centering
\includegraphics[width=0.7\textwidth]{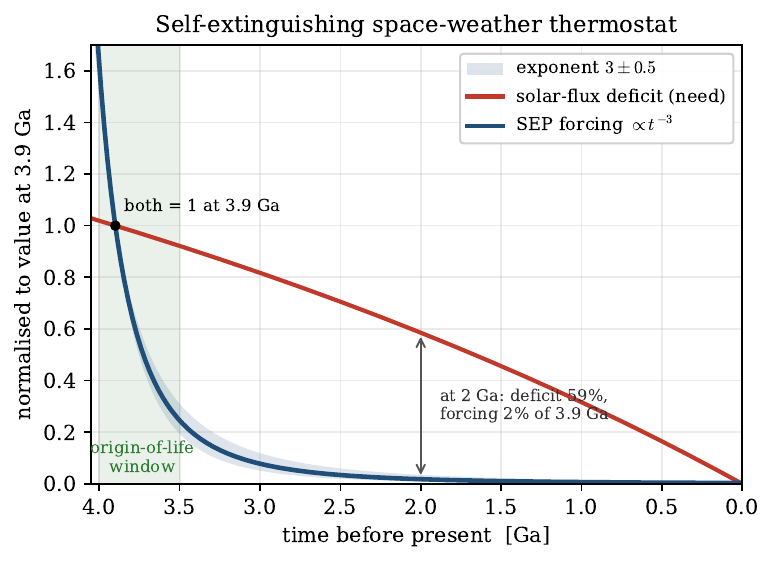}
\caption{The self-extinguishing thermostat, shown as normalized time-shapes. Both curves are
divided by their values at 3.9~Ga, so the figure compares how fast each changes;
\S\ref{sec:netforcing} gives their magnitudes. The solar-flux deficit (red), from the
standard luminosity history \citep{feulner2012}, shrinks slowly as the Sun brightens and is still
$\sim$60\% of its 3.9~Ga value at 2~Ga. Over the same interval the deceleration-powered SEP
forcing (blue, $\propto t^{-3}$; shaded band for exponents 2.5--3.5) falls to $\sim$2\%. The
forcing is therefore a Hadean/Eoarchean booster that switches off before it can overheat the
later Earth. The two curves change for independent physical reasons (core hydrogen burning for
the luminosity, magnetic braking for the activity), and both follow stellar age, so they are
anti-correlated over the main sequence.}
\label{fig:thermostat}
\end{figure}

\subsection{A self-extinguishing thermostat tied to the wind history}
\label{sec:thermostat}

Because the SEPs are powered by deceleration, the forcing inherits the measured wind-density
history. The deceleration power scales as
\begin{equation}
  P_{\mathrm{decel}} \propto n_{\mathrm{amb}}\,f_{\mathrm{CME}}\,\Delta E_k .
\end{equation}
With Skumanich spin-down \citep[$v_{\mathrm{rot}}\propto t^{-1/2}$;][]{skumanich1972}, Wood mass loss
\citep[$n_{\mathrm{amb}}\propto t^{-2.3}$;][]{wood2005}, and CME rate and per-event energy both declining with
activity, the net is $P_{\mathrm{decel}}\propto t^{-3}$ or steeper. Over stellar age
$0.6\to2.5\unit{Gyr}$ (i.e.\ $3.9\to2\unit{Ga}$),
\begin{equation}
  \frac{P_{\mathrm{decel}}(3.9\,\mathrm{Ga})}{P_{\mathrm{decel}}(2\,\mathrm{Ga})}
  \sim \left(\frac{2.5}{0.6}\right)^{3} \approx 10^{2}.
\end{equation}
The SEP-greenhouse forcing was $\sim 100\times$ stronger when the Sun was faintest and faded
two orders of magnitude as it brightened $\sim 12\%$. Extrapolated to the present solar age of
4.57~Gyr, the same scaling gives $(4.57/0.6)^3\approx4\times10^{2}$: the forcing at 3.9~Ga was
several hundred times its present-day level. The forcing and the luminosity are anti-correlated
even though different processes drive them. The luminosity rises because core hydrogen burning
raises the mean molecular weight of the core, which contracts and heats. The forcing declines
because magnetic braking spins the Sun down and weakens its activity and wind. Both processes
proceed monotonically with stellar age over the main sequence, in opposite senses, so the
anti-correlation is guaranteed. The rate is the informative part: the forcing falls as
$\sim t^{-3}$, while the deficit shrinks by only $\sim$40\% between 3.9 and 2~Ga
(Fig.~\ref{fig:thermostat}). The measured wind history, with no free parameter, sets when the
forcing switches off.

\subsection{Spectral hardness and the sign of the forcing; intermittency}

The same events drive a nitrogen channel ($\mathrm{N_2O}$, greenhouse) and a carbon channel
(HCN $\to$ organic haze, generally anti-greenhouse). The deposition altitude, and with it the sign
of the net forcing, is set by $\gamma$: hard spectra favour deep, long-lived $\mathrm{N_2O}$
(warming); soft spectra favour a high haze (cooling). The haze in turn shields $\mathrm{N_2O}$
(and any $\mathrm{NH_3}$, $\mathrm{CH_4}$) from UV photolysis, lengthening lifetimes and raising
the amplification. Because $\tau_{\mathrm{N_2O}}\gg\Delta t_{\mathrm{event}}$ (decades versus days), the
reservoir smooths the bursty production into a quasi-steady forcing. UV loss also scales with
activity, which partly offsets production and buffers the steady-state inventory. Mean-forcing
climate models are therefore adequate, and individual superflares do not destabilise the
warming.

\subsection{Position relative to other FYS resolutions}

As a source of magnitude, SEP-driven $\mathrm{N_2O}$ contributes a few $\unit{W\,m^{-2}}$
(\S\ref{sec:netforcing}). This is comparable to or below the $\mathrm{CO_2}$, $\mathrm{CH_4}$, and
collision-induced channels \citep{kasting1993,wordsworth2013}, and it adds to them. Its distinctive
contribution lies in its time-dependence and in the evidence it leaves. Every greenhouse
resolution must explain a warm Hadean and also the absence of runaway warmth once the Sun
brightened. $\mathrm{CO_2}$--$\mathrm{CH_4}$ scenarios achieve the second through carbon-cycle or
biological feedbacks imposed externally. The deceleration-powered forcing instead declines as
$\sim t^{-3}$ on its own, because it follows the Sun's rotational spin-down. The luminosity it
offsets rises for an independent reason, core hydrogen burning, and far more slowly
(\S\ref{sec:thermostat}). The SEP forcing is therefore best regarded as a self-extinguishing
booster that works with the conventional greenhouse during the faint-Sun epoch and then
disappears. Through its nitrogen-isotope signature (\S\ref{sec:synthesis}), it is also one of the
few FYS contributors that can be tested independently in the rock record.

A recent quantitative test sharpens this position. \citet{biasiotti2026} coupled a
thermochemical--photochemical kinetics model to a radiative--convective and energy-balance climate
chain and drove an Eoarchean atmosphere with a sustained Carrington-class SEP flux. They found
that greenhouse species produced by space weather warm the surface by $\lesssim0.3\unit{K}$, with
$\mathrm{N_2O}$ and HCN negligible and CO and $\mathrm{H_2}$ dominating the small residual. On this
basis they exclude SEP-driven $\mathrm{N_2O}$ as a stand-alone FYS solution, contrary to
\citet{airapetian2016}. Two features of their setup place the present mechanism in a regime they
did not probe. First, they compute the steady-state processed composition under a continuous flux
and defer the time-dependent chemistry of transient events. Our forcing is a transient,
self-extinguishing pulse (\S\ref{sec:thermostat}), and its signature is its time structure.
Second, they exclude haze-forming mixtures ($\mathrm{CH_4}/\mathrm{CO_2}\gtrsim0.1$) and find that
the SEP-produced $\mathrm{N_2O}$ forms high in the column, where its integrated abundance and
radiative effect are small. In the hazy regime of our net-forcing balance
(\S\ref{sec:netforcing}), a fractal haze attenuates the UV that would otherwise photolyse that
high-altitude $\mathrm{N_2O}$ \citep{wolftoon2010} and partly preserves it (\S\ref{sec:uvdecouple}).
We therefore read \citet{biasiotti2026} as bounding the haze-free, steady-state $\mathrm{N_2O}$
contribution to near zero, while the transient, haze-dependent channel developed here remains
open. We accordingly keep $\mathrm{N_2O}$ as a contributing term in the budget.

\section{Role in the origin of life}
\label{sec:ool}

Many settings have been proposed for the origin of life, and the question remains actively
debated. For the purposes of this paper we focus on three that differ in their access to
atmospherically delivered feedstock: submarine hydrothermal vents, ocean tide pools, and
freshwater (subaerial) thermal pools. The submarine hydrothermal-vent hypothesis \citep{baross1985,russell1997,martin2008,sojo2016}
emphasises the sustained redox and pH gradients of hydrothermal vents and their
continuity with chemiosmotic metabolism and the inferred autotrophy of the last universal common
ancestor. The subaerial hot-spring hypothesis \citep{mulkidjanian2012,damer2020} rests on two main
arguments. Condensation of nucleic acids and peptides is thermodynamically unfavourable in bulk
water without some change in water activity and requires wet--dry cycling \citep{rossdeamer2016}.
The high-$\mathrm{K^+}$, high-phosphate, low-$\mathrm{Na^+}$ composition of cytoplasm also
resembles vapour-dominated geothermal fields more than seawater \citep{mulkidjanian2012}. Recent
experiments support the subaerial route: wet--dry cycling drives nucleotide oligomerisation into
long chains without added activators \citep{song2024,caimi2025}. The supply of feedstock connects
to space weather. SEP irradiation of weakly reducing atmospheres could yield amino and carboxylic
acids \citep{kobayashi2023} and HCN, the central input to cyanosulfidic prebiotic chemistry
\citep{powner2009,patel2015}. This links the origin of life to stellar activity and to the
atmosphere above the surface environments where it may have occurred
\citep{sasselov2020,benner2020}. Using the SEP fluxes of \S\ref{sec:sepprod}, we now quantify feedstock delivery to these three
settings.

\subsection{Feedstock and delivery to the surface}
\label{sec:feedstock}

SEP precipitation into an $\mathrm{N_2}$--$\mathrm{CO_2}$--$\mathrm{CH_4}$ atmosphere fixes
nitrogen (NO, $\mathrm{NO_2}$, $\mathrm{N_2O}$ $\to$ nitrate/nitrite) and, through the carbon
channel, produces HCN. Spreading the global
rate of Eq.~\eqref{eq:globalN} over Earth's surface gives an areal deposition flux
\begin{equation}
  \phi \sim \frac{1.4\e{10}\text{--}1.4\e{11}\unit{mol\,yr^{-1}}}{5.1\e{14}\unit{m^2}}
  \sim 3\e{-5}\text{--}3\e{-4}\unit{mol\,m^{-2}\,yr^{-1}} .
\end{equation}
For a pool of depth $d\sim 0.3\unit{m}$, bare in-fall raises the dissolved fixed-N concentration
by
\begin{equation}
  \frac{\Delta c}{\Delta t} = \frac{\phi}{d}
  \sim \frac{10^{-4}}{0.3} \sim 3\e{-4}\unit{mol\,m^{-3}\,yr^{-1}}
  \approx 0.3\,\mu\mathrm{M\,yr^{-1}} .
\end{equation}
Two factors raise this in a real geothermal field: a catchment larger than the pool surface
($\sim 10\text{--}100\times$) and wet--dry cycling ($\sim 10\text{--}1000\times$ before reset).
Together they can take a sub-$\mu$M signal to a steady-state inventory of order $\mu$M--mM, the
range that prebiotic nitrogen and cyanosulfidic chemistry require.

\subsection{Comparison of candidate environments}

The vent setting is energetically self-contained and shielded, which is attractive under a
hostile young Sun, but it receives the atmospheric feedstock the CMEs produce only after dilution through the ocean
column. Tide
pools offer wet--dry cycling but inherit the chemistry of seawater: divalent cations disrupt
fatty-acid vesicle self-assembly and precipitate phosphate. Of the three, freshwater thermal pools
are the only setting that receives both: SEP-fixed nitrogen and HCN from above, and low-salt,
high-$\mathrm{K^+}$/phosphate, wet--dry chemistry locally.

\begin{table}[ht]
\centering
\caption{How the SEP feedstock maps onto the three candidate settings.}
\label{tab:envs}
\small
\begin{tabular}{@{}llll@{}}
\toprule
\cell{0.15}{} & \cell{0.22}{\textbf{Submarine vents}} & \cell{0.22}{\textbf{Ocean tide pools}} &
\cell{0.25}{\textbf{Freshwater thermal pools}} \\
\midrule
\cell{0.15}{SEP feedstock access} &
\cell{0.22}{None: km of seawater shields all SEP/UV; product diluted to $\sim 10^{-10}\unit{M}$ in the ocean column} &
\cell{0.22}{Direct rain-out, but into seawater} &
\cell{0.25}{Direct rain-out into the pool and its catchment} \\[2pt]
\cell{0.15}{Wet--dry cycling} &
\cell{0.22}{No classic wet--dry cycling (submerged); changes in water activity possible} &
\cell{0.22}{Yes} &
\cell{0.25}{Yes} \\[2pt]
\cell{0.15}{Ionic chemistry} &
\cell{0.22}{Internal redox (serpentinisation $\mathrm{H_2}$); decoupled from the Sun} &
\cell{0.22}{High $\mathrm{Na^+}$; $\mathrm{Mg^{2+}/Ca^{2+}}$ disrupt vesicles and precipitate phosphate} &
\cell{0.25}{High $\mathrm{K^+}$/phosphate, low $\mathrm{Na^+}$: matches cytoplasm} \\[2pt]
\cell{0.15}{Summary} &
\cell{0.22}{Radiation refuge; atmospheric feedstock arrives only diluted} &
\cell{0.22}{Wet--dry geometry, but seawater chemistry} &
\cell{0.25}{Receives feedstock and supports the downstream chemistry} \\
\bottomrule
\end{tabular}
\end{table}

\subsection{Prebiotic chemistry: why the surface, and a thermal caveat}

Polymerisation of nucleic acids and peptides is thermodynamically unfavourable in bulk water with
high water activity, because hydrolysis dominates condensation. Wet--dry cycling, or more
generally cycling of water activity, supplies the activation by repeatedly concentrating
reactants at air--water interfaces. At a submerged vent, changes in water activity remain
possible, but classic wet--dry cycling is not. The
ionic-composition argument independently favours freshwater geothermal fields: the cytoplasm of
all cells is high-$\mathrm{K^+}$, high-phosphate, low-$\mathrm{Na^+}$, which resembles
vapour-dominated geothermal fields more than seawater.

One caveat applies to hot pools specifically. HCN hydrolyses in minutes to hours at
$70\text{--}90^\circ$C and near-neutral pH, so a boiling pool destroys cyanide as fast as it
arrives. Nitrate and nitrite are stable at these temperatures, while HCN preservation favours
cooler margins, eutectic freezing, or ferrocyanide sequestration. The simplest setting that meets
both constraints is a thermally graded field, with hot zones for polymerisation and cool catchment
pools for cyanide accumulation. Both are fed by the same SEP rain-out, which redistributes
globally before deposition and so reaches mid-latitude pools despite the polar-weighted
precipitation (Table~\ref{tab:envs}).

\section{Synthesis, testable predictions, and extensions}
\label{sec:synthesis}

Stellar activity affects both climate and the prebiotic inventory, and ``space-weather
habitability'' is now an active research area \citep{airapetian2020}. For the early Earth, the
energetic-particle forcing that fixes nitrogen should leave a geochemical record. The Archean
nitrogen-isotope archive constrains the timing and pathways of nitrogen fixation
\citep{stueken2016} and so offers, in principle, a test for an abiotic SEP source distinct from
biological and lightning fixation. For planets orbiting M~dwarfs, frequent flares and dense winds
intensify the same forcing, with competing consequences for atmospheric chemistry and ozone
shielding \citep{segura2010,tilley2019}, while the surface ultraviolet environment governs whether
cyanosulfidic chemistry can proceed at all \citep{ranjan2016,rimmer2018}. We therefore treat the
climatic and prebiotic consequences together, as products of one spectrum-dependent forcing that
fades with stellar age, and list the predictions that follow.

\subsection{Uncertainty budget}
\label{sec:robust}

Because the argument is a chain, we separate the secure links from those that carry the
uncertainty. Table~\ref{tab:robust} classifies each quantity as model-independent (fixed by
geometry or endpoints), model-dependent (sensitive to a propagation or transport model), or
assumed (adopted from modern calibration or extrapolation). The energetic anchor $\Delta E_k$ and
the qualitative time-dependence $\propto t^{-3}$ are secure, whereas the absolute amplitude of
every downstream consequence is uncertain. Most of the uncertainty sits in the spectral break
energy $E_b$, the event frequency $\nu$, and the $\mathrm{N_2O}$ lifetime $\tau$. Closing the FYS
budget with $\mathrm{N_2O}$ would need all three at their favourable ends at once, which is why we
treat $\mathrm{N_2O}$ as a contributing term. For the nitrogen yield, the survival pressure $p_s$
adds a factor of $\sim$3 either way. The shape of the forcing in time follows from the wind
history alone and so holds across the full uncertainty range.

\begin{table}[h]
\centering
\caption{Uncertainty budget for the principal quantities. ``Class'' is model-independent (MI;
fixed by endpoints or geometry), model-dependent (MD; sensitive to a propagation/transport
model), or assumed (A; from modern calibration or extrapolation).}
\label{tab:robust}
\small
\begin{tabular}{@{}llll@{}}
\toprule
Quantity & Nominal & Class & Dominant uncertainty (range) \\
\midrule
$\Delta E_k$ (deceleration energy)      & $6\e{25}\unit{J}$        & MI & CME mass, $v_0$ ($\pm0.5$~dex) \\
$\varepsilon$ (DSA efficiency)          & $0.1$                    & A  & shock efficiency ($\times2$--3) \\
$\Esep$                                 & $6\e{24}\unit{J}$        & MI/derived & via $\varepsilon$ ($\pm0.5$~dex) \\
$b$ (beaming factor)                    & $\approx6$               & MD & shock angular extent ($\times1.5$) \\
$E_b$ (spectral break)                  & $30$--$100\unit{MeV}$    & MD & diffusion ($\eta_{\rm Bohm}$; factor $\sim3$) \\
$\fdep$ (deposited fraction)            & $0.15$                   & MD & via $E_b$, $\gamma_2$, $p_s$ ($0.04$--$0.7$) \\
$p_s$ (survival pressure)               & $1\unit{hPa}$            & A  & photochemical survival of fixed N ($0.15$--$3\unit{hPa}$) \\
$G$ (fixed N per joule)                 & $1.2\e{17}\unit{J^{-1}}$ & A  & excited-state fraction; anoxic chemistry ($\times2$) \\
$\nu$ (event frequency)                 & $\sim10^{2}\unit{yr^{-1}}$ & A & superflare statistics ($\pm1$--$1.5$~dex) \\
$f_{\rm acc}$ (magnetospheric access)   & $\gtrsim1$               & A  & early geomagnetic field (factor of a few, $\geq1$) \\
$r_A$ (Alfv\'en radius)                 & $40$--$60\,\Rsun$        & MD & mass-loss history (factor $\sim1.5$) \\
$\tau_{\mathrm{N_2O}}$ (lifetime)       & $50$--$5000\unit{yr}$    & A  & haze shielding/morphology ($\times10$--100) \\
$\Delta F_{\rm net}$ (net N$_2$O forcing) & few $\unit{W\,m^{-2}}$ & derived & haze morphology, $\tau$ (sign-dependent) \\
$P_{\rm decel}\propto t^{-3}$ (time-shape) & $\cdots$             & MI & wind/spin-down history (exponent $\pm0.5$) \\
\bottomrule
\end{tabular}
\end{table}

\subsection{Climate and prebiotic consequences of one forcing}

SEPs from decelerating young-Sun CMEs warm the planet through the $\mathrm{N_2O}$ greenhouse
(\S\ref{sec:fys}) and stock surface pools with fixed nitrogen and HCN (\S\ref{sec:ool}). Both
outputs scale with the same declining activity ($\propto t^{-3}$), so both peak in the Hadean and
early Archean, when life arose, and fade together as the Sun spins down. A warm early Earth and a
chemically primed early Earth are then two outcomes of one space-weather forcing, each gated by
the same break energy $E_b$, and need not be treated as independent coincidences.

\begin{figure}[h]
\centering
\includegraphics[width=0.8\textwidth]{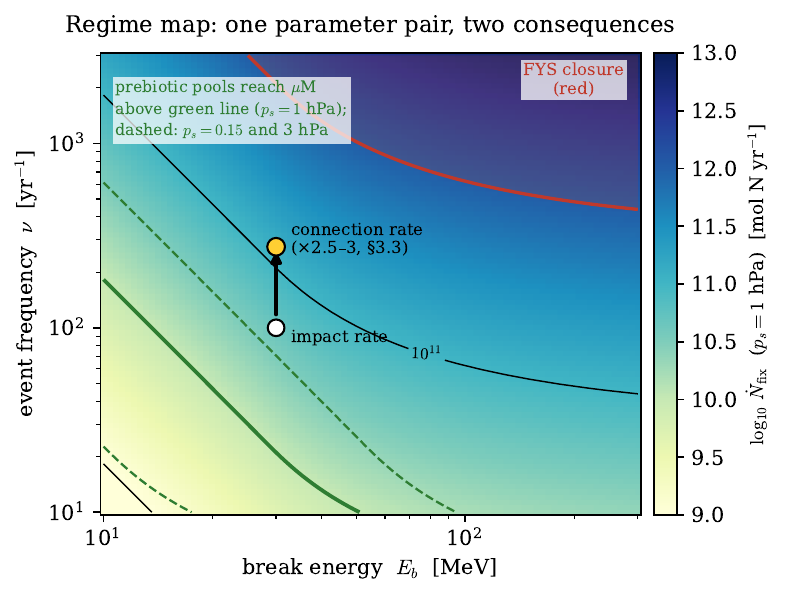}
\caption{Regime map in the (break energy $E_b$, event frequency $\nu$) plane. Colour and
contours give the global fixation rate $\dot N_{\rm fix}=3.3\e{9}\,\fdep(p_s;E_b)\,\nu$ for the
nominal $p_s=1$~hPa. The solid green line marks where geothermal pools reach $\mu$M feedstock
($\dot N_{\rm fix}=10^{10}\unit{mol\,N\,yr^{-1}}$); dashed green lines show the same threshold for
$p_s=0.15$ and 3~hPa. The red region marks FYS closure ($\dot N_{\rm fix}\ge10^{12}\unit{mol\,N\,
yr^{-1}}$, requiring $\mathrm{N_2O}$ lifetime extension). The nominal operating point lies inside
the prebiotic region for every survival level, at both the impact and the connection rates, but
nearly an order of magnitude below FYS closure: the same parameter pair
drives both consequences, with climate far more demanding.}
\label{fig:regime}
\end{figure}

\subsection{Testable predictions}

\begin{enumerate}
  \item \textit{The break energy $E_b$.} Both a meaningful climatic contribution
  (\S\ref{sec:budget}) and adequate prebiotic concentrations (\S\ref{sec:feedstock}) require a
  penetrating break, $E_b\gtrsim30\unit{MeV}$. This is equivalent to a hard low-energy slope
  $\gamma_1\lesssim2$ with the rollover above the survival threshold $E_s$ ($\approx$28~MeV for
  $p_s=1$~hPa; Figs.~\ref{fig:fpen} and \ref{fig:regime}). It can be constrained by stellar SEP
  proxies from EK~Dra--type eruptions on young solar analogues \citep{namekata2025} and by
  EUHFORIA inner-heliosphere runs that output the shock Mach-number history $X(r)$ along the
  transit, from which the event-integrated $\gamma$ follows.
  \item \textit{Nitrogen-isotope signature.} SEP-driven fixation should imprint a distinctive
  $\delta^{15}\mathrm{N}$ (and possibly mass-independent) signature on Archean nitrate/ammonium
  preserved in sediments, separable from biological and lightning fixation. A systematic
  $\delta^{15}\mathrm{N}$ excursion tracking the predicted $t^{-3}$ activity decline would be
  strong support.
  \item \textit{Latitude structure.} Polar-cap-weighted precipitation predicts a
  palaeo-latitudinal gradient in fixed-N delivery, partly smoothed by stratospheric transport
  before rain-out. The gradient could in principle be tested against the geographic distribution
  of the earliest biosignatures.
\end{enumerate}

\subsection{Extension to Mars and exoplanets}

The mechanism is generic to magnetically active stars. A weakly magnetised early Mars would receive the same SEP-driven fixation, which bears on its own
faint-Sun warming and prebiotic inventory. Because the early Martian atmosphere collapsed, any
coevolution of life and environment there would have followed a different path from Earth's
\citep{cabrol2018}, and the decline of the SEP forcing with time is one input to that history. For M-dwarf planets, frequent flares and dense winds make the deceleration channel
even stronger, but the same $\gamma$-gating applies: only hard event-integrated spectra deliver
the deep, surviving chemistry. The framework therefore turns the statement that a star is active
into a quantitative, spectrum-dependent habitability criterion.

\subsection*{Summary of the chain}

\begin{center}
\fbox{\parbox{0.95\textwidth}{\centering
$\Delta E_k\,(6\e{25}\,\mathrm{J})
\;\xrightarrow{\varepsilon\sim0.1}\;
\Esep\,(6\e{24}\,\mathrm{J})
\;\xrightarrow{b\approx6}\;
E_{\mathrm{atm}}\,(1.6\e{16}\,\mathrm{J})
\;\xrightarrow{\fdep(E_b,\,p_s)}\;
E_{\mathrm{dep}}
\;\xrightarrow{G}\;
\dot N_{\mathrm{fix}}$\\[4pt]
$\Longrightarrow$ \; greenhouse forcing (FYS) \; and \; surface feedstock (origin of
life), \; both $\propto t^{-3}$, \; both gated by $E_b$.}}
\end{center}

\section{Conclusions}
\label{sec:conclusions}
 
We have built a semi-analytic model chain that follows an eruption on the young Sun to a
climatic and prebiotic forcing of the early Earth.

A fast CME ($v_0\sim3\text{--}4\e{3}\unit{km\,s^{-1}}$) decelerating in the dense inner
heliosphere dissipates $\Delta E_k\sim6\e{25}\unit{J}$. This energy is set by the launch and final
speeds alone, so it does not depend on the propagation model. Diffusive shock acceleration turns
a fraction of it into SEPs, and the depth at which they deposit energy depends mainly on the
spectral break energy $E_b$. The stronger young-Sun field moves the Alfv\'en surface out to
$\sim$40--60$\,\Rsun$ (\S\ref{sec:alfven}), and the CME-shock channel then delivers more than ten
times the particle fluence of the flare channel (\S\ref{sec:transport}). The strong field also
raises the coronal Alfv\'en speed, so a CME must reach $\sim$3500$\unit{km\,s^{-1}}$ to escape.

The forcing declines as $\sim t^{-3}$ because it is powered by deceleration and so inherits the
measured $\dot M\propto t^{-2.3}$ wind history \citep{wood2005}. The solar luminosity rises for a
different reason, core hydrogen burning, and both trends are monotonic in stellar age. The
forcing was therefore strongest when the Sun was faintest and weakened as it brightened
(\S\ref{sec:thermostat}). Greenhouse resolutions of the faint young Sun problem otherwise have to
build in such a decline through carbon-cycle or biological feedbacks. This time structure depends
only on the spin-down history and holds across the full parameter uncertainty
(Table~\ref{tab:robust}).

The magnitude of the forcing is small and depends on parameters. SEP-driven $\mathrm{N_2O}$ gives
a net positive forcing of a few $\unit{W\,m^{-2}}$ only when the haze is fractal; for compact
spherical haze the net is negative (\S\ref{sec:netforcing}). This agrees with the steady-state,
haze-free bound of $\lesssim0.3\unit{K}$ from \citet{biasiotti2026}, in which CO and
$\mathrm{H_2}$ carry the small residual. We therefore treat $\mathrm{N_2O}$ as a contributing term
in a transient, haze-dependent regime. The same particles fix
$\sim$$4\e{10}$--$4\e{11}\unit{mol\,N\,yr^{-1}}$ below a survival level near 1~hPa and could stock
surface pools with fixed nitrogen and HCN (\S\ref{sec:ool}). The climatic and prebiotic
consequences share one driver, and both scale as $t^{-3}$.

The scheme can be tested. A meaningful climatic contribution and adequate prebiotic
concentrations both require $E_b\gtrsim30\unit{MeV}$ (Fig.~\ref{fig:regime}). Stellar SEP proxies
on young solar analogues such as EK~Dra \citep{namekata2025} and inner-heliosphere simulations of
the shock Mach-number history can constrain this. SEP-driven fixation should also leave a
$\delta^{15}\mathrm{N}$ signature in Archean sediments that tracks the $t^{-3}$ decline, together
with a palaeo-latitudinal gradient in fixed-nitrogen delivery (\S\ref{sec:synthesis}). The next
steps are to couple the SEP flux to a co-evolving palaeo-magnetospheric model and to replace our
band model with a coupled photochemical--radiative calculation. That calculation would also fix
the survival pressure $p_s$ and treat the $\mathrm{N_2O}$--haze balance and the particle--UV
decoupling explicitly.

The mechanism applies to any magnetically active star. It predicts that the strongest
space-weather forcing coincides with the faint-star epoch and then fades, for the early Earth and
early Mars as well as for planets orbiting active M~dwarfs.

\begin{acknowledgements}

\end{acknowledgements}

\software{NumPy \citep{harris2020}, Matplotlib \citep{hunter2007}}

\end{document}